\documentclass[
    reprint, 
    amsmath, 
    amssymb, 
    aps, 
    prd, 
    superscriptaddress,
    nofootinbib,
    ]{revtex4-2}

\usepackage{graphicx}

\usepackage{hyperref}

\usepackage{physics}

\usepackage[caption=false]{subfig}

\usepackage{orcidlink}

\usepackage[10pt]{type1ec} 
\usepackage[T1]{fontenc}
\usepackage{CJKutf8}
\usepackage[overlap, CJK]{ruby}
\usepackage{CJKulem}
\newenvironment{SChinese}{%
  \CJKfamily{gbsn}%
  \CJKtilde
  \CJKnospace}{}

\newcommand{\nTOV}{n_\mathrm{TOV}}
\newcommand{\MTOV}{M_\mathrm{TOV}}
\newcommand{\nsat}{n_\mathrm{sat}}
\newcommand{\muH}{\mu_\mathrm{H}}
\newcommand{\muT}{\mu_\mathrm{T}}
\newcommand{\muL}{\mu_\mathrm{L}}
\newcommand{\nH}{n_\mathrm{H}}
\newcommand{\nT}{n_\mathrm{T}}
\newcommand{\nL}{n_\mathrm{L}}
\newcommand{\pH}{p_\mathrm{H}}
\newcommand{\pT}{p_\mathrm{T}}
\newcommand{\pL}{p_\mathrm{L}}
\newcommand{\csq}{c^2_{s}}

\newcommand{\CIT}{\affiliation{TAPIR, California Institute of Technology, Pasadena, CA 91125, USA}}
\newcommand{\CITLab}{\affiliation{LIGO Laboratory, California Institute of Technology, Pasadena, CA 91125, USA}}
\newcommand{\CITA}{\affiliation{Canadian Institute for Theoretical Astrophysics, Toronto, ON M5S 3H8, Canada}}
\newcommand{\UOT}{\affiliation{Department of Physics, University of Toronto, Toronto, ON M5S 1A7, Canada}}
\newcommand{\DADDAA}{\affiliation{David A. Dunlap Department of Astronomy \& Astrophysics, University of Toronto, Toronto, ON M5S 3H4, Canada}}
\newcommand{\PI}{\affiliation{Perimeter Institute for Theoretical Physics, Waterloo, ON N2L 2Y5, Canada}}
\newcommand{\TDLI}{\affiliation{Tsung-Dao Lee Institute, Shanghai, 201210, China}}
\newcommand{\SJTU}{\affiliation{School of Physics and Astronomy, Shanghai Jiao Tong University, Shanghai, 200240, China}}

\begin{document}

\title{Unified nonparametric equation-of-state inference from the neutron-star crust to perturbative-QCD densities}

\author{Eliot Finch 
\orcidlink{0000-0002-1993-4263}}
\email{efinch@caltech.edu}
\CIT

\author{Isaac Legred
\orcidlink{0000-0002-9523-9617}}
\email{ilegred@caltech.edu}
\CIT
\CITLab

\author{Katerina Chatziioannou
\orcidlink{0000-0002-5833-413X}}
\email{kchatziioannou@caltech.edu}
\CIT
\CITLab

\author{Reed Essick
\orcidlink{0000-0001-8196-9267}}
\email{essick@cita.utoronto.ca}
\CITA
\UOT
\DADDAA

\author{Sophia Han 
(\begin{CJK}{UTF8}{}\begin{SChinese}韩 君\end{SChinese}\end{CJK})
\orcidlink{0000-0002-9176-4617}
}
\email{sjhan@sjtu.edu.cn}
\TDLI
\SJTU
\affiliation{State Key Laboratory of Dark Matter Physics, Shanghai Jiao Tong University, Shanghai 201210, China}

\author{Philippe Landry
\orcidlink{0000-0002-8457-1964}}
\email{pgjlandry@gmail.com}
\CITA
\PI

\date{\today}

\begin{abstract}

Perturbative quantum chromodynamics (pQCD), while valid only at densities exceeding those found in the cores of neutron stars, could provide constraints on the dense-matter equation of state (EOS). 
In this work, we examine the impact of pQCD information on the inference of the EOS using a nonparametric framework based on Gaussian processes (GPs). 
We examine the application of pQCD constraints through a ``pQCD likelihood,'' and verify the findings of previous works; namely, a softening of the EOS at the central densities of the most massive neutron stars and a reduction in the maximum neutron-star mass. 
Although the pQCD likelihood can be easily integrated into existing EOS inference frameworks, this approach requires an arbitrary selection of the density at which the constraints are applied.
The EOS behavior is also treated differently on either side of the chosen density.
To mitigate these issues, we extend the EOS model to higher densities, thereby constructing a ``unified'' description of the EOS from the neutron-star crust to densities relevant for pQCD. 
In this approach the pQCD constraints effectively become part of the prior. 
Since the EOS is unconstrained by any calculation or data between the densities applicable to neutron stars and pQCD, we argue for maximum modeling flexibility in that regime.
We compare the unified EOS with the traditional pQCD likelihood, and although we confirm the EOS softening, we do not see a reduction in the maximum neutron-star mass or any impact on macroscopic observables.
Though residual model dependence cannot be ruled out, we find that pQCD suggests the speed of sound in the densest neutron-star cores has already started decreasing toward the asymptotic limit; we find that the speed of sound squared at the center of the most massive neutron star has an upper bound of $\sim 0.5$ at the 90\% level.
\end{abstract}

\maketitle

\section{Introduction}
\label{sec:introduction}

The cold dense-matter equation of state (EOS) within neutron stars (NSs)~\cite{Baym:2017whm} encodes macroscopic NS properties~\cite{Chatziioannou:2024tjq} such as the mass and radius via the Tolman-Oppenheimer-Volkoff (TOV) equations~\cite{Tolman:1939jz, Oppenheimer:1939ne} and the tidal deformability~\cite{Hinderer:2007mb}.
The EOS, a relation between the pressure and energy density, is in principle determined by the underlying theory of strong interactions, quantum chromodynamics (QCD).
However, in practice, calculations are only possible in limiting cases at low and high densities.
At low densities, roughly at or below nuclear saturation density ($\nsat \approx 0.16\,\mathrm{fm}^{-3}$, or $\nsat m_u \approx 2.7 \times 10^{14}\,\mathrm{g}\,\mathrm{cm}^{-3}$ where $m_u$ is the atomic mass constant), quarks are confined to hadrons and interactions can be described with effective theories of QCD~\cite{Hammer:2019poc}, for example chiral effective field theory ($\chi$EFT)~\cite{Drischler:2021kxf}.
This density regime applies to the outermost layers of a NS, typically to depths of $\sim 1\,\mathrm{km}$~\cite{Haensel:2007yy, Baym:2017whm}.

Densities increase toward the core of the NSs, reaching $\lesssim 6\,\nsat$~\cite{Lattimer:2004pg, Ozel:2016oaf}.
However, beyond $\sim \nsat$ calculations of the dense-matter EOS become intractable and thus much of our knowledge comes from astronomical observations.
The masses of numerous NSs have been obtained via radio observations~\cite{Ozel:2016oaf}, the most massive of which~\cite{NANOGrav:2019jur,Fonseca:2021wxt} provide a lower limit on the maximum mass possible and thus the EOS at $\sim 6\,\nsat$~\cite{Ozel:2016oaf}.
X-ray hot-spot modeling via NICER~\cite{Watts:2016uzu} allows for joint mass and radius constraints~\cite{Baym:2017whm}. 
Though not considered here, X-ray observations can lead to further mass and radius constraints via spectroscopic measurements~\cite{Ozel:2016oaf}.
Finally, gravitational-wave observations of binary NS mergers yield measurements of NS masses and tidal deformabilities~\cite{Chatziioannou:2020pqz}.
There are currently two such observations in the catalog of events produced by the LIGO-Virgo-KAGRA network~\cite{LIGOScientific:2014pky,VIRGO:2014yos,KAGRA:2020tym}.

For densities beyond those encountered in NSs, namely $\sim 6\,\nsat$, astronomical observations have little constraining power. 
It is not until much higher densities of $\sim 40\,\nsat$ that nuclear calculations become possible again.
At increasingly high densities, QCD becomes asymptotically free and a phase transition to deconfined quark matter occurs. 
The EOS can then be calculated via perturbative QCD (pQCD)~\cite{Kurkela:2009gj}. 
Although describing the EOS only at densities much beyond NS densities, pQCD could still provide constraints on the NS EOS since a single, self-consistent EOS must span from NS to pQCD densities.

Quantifying the impact of pQCD on the properties of NSs hinges on assumptions about how to treat the $(6 \text{--} 40)\,\nsat$ regime.
One of the first attempts to incorporate pQCD into EOS modeling was performed by \citet{Kurkela:2014vha} (see also \citet{Fraga:2015xha, Annala:2017llu}), who employed piecewise polytropes to extend the EOS to pQCD densities.
Since this initial work, a variety of methods have been used to bridge the low- and high-density regimes; these include a ``speed-of-sound'' interpolation~\cite{Annala:2019puf,Annala:2021gom,Altiparmak:2022bke}, connecting relativistic mean-field and MIT-Bag models~\cite{Shirke:2022tta}, and Bayesian model mixing~\cite{Semposki:2024vnp, Semposki:2025etb}.

Recently, instead of attempting to explicitly model the full range of densities, pQCD constraints have been incorporated into existing EOS models via a ``pQCD likelihood.''
\citet{Komoltsev:2021jzg} derived analytical constraints on the EOS at NS densities based on the required behavior at pQCD densities and provided a likelihood function that can be straightforwardly applied to candidate EOSs~\cite{komoltsev_2023_7781233}.
Perhaps in part due to its ease of use, this approach has been used in a number of works~\cite{Gorda:2022jvk,Somasundaram:2022ztm,Han:2022rug,Gorda:2023usm,Annala:2023cwx,Zhou:2023zrm,Mroczek:2023zxo,Fan:2023spm}, including \citet{Gorda:2022lsk} who also use the EOS draws of \citet{Annala:2021gom} to check the consistency of candidate EOS with pQCD.
This approach of using a separate EOS model (known to be consistent with pQCD) to judge whether low-density candidate EOSs are consistent with pQCD foreshadowed the introduction of a new pQCD likelihood function based on nonparametric EOS extensions in work by~\citet{Komoltsev:2023zor}.
Although some of the above studies reach different quantitative conclusions about the impact of pQCD (as discussed in \citet{Komoltsev:2023zor}), the consensus is a general softening of the EOS at the cores of the most massive NSs close to the TOV limit and a corresponding reduction in the maximum TOV NS mass.

In this work we revisit the pQCD constraints. 
After a pedagogical overview of the pQCD constraints on cold dense matter in Sec.~\ref{sec:pqcd_constraints}, we perform a detailed exploration of the various proposed ``pQCD likelihoods'' on a highly flexible, nonparametric EOS model~\cite{Landry:2018prl, Essick:2019ldf, Landry:2020vaw}. 
We find broad qualitative agreement with previous results.
Along the way, we clarify the various assumptions and limitations of each---see Sec.~\ref{sec:post-hoc}.
Then, in Sec.~\ref{sec:pqcd_prior}, we return to an approach more akin to earlier works which model the full range of densities; we construct a flexible ``unified'' nonparametric EOS model that satisfies pQCD as part of the prior and thus describes nuclear matter from the NS crust to densities relevant to pQCD.
This approach avoids some of the drawbacks associated with the pQCD likelihoods, while striving for model flexibility.
Since the EOS is unconstrained between NS central densities and pQCD, we aim for maximum model flexibility in this regime.
Under this flexible unified EOS model, we find that pQCD offers informative constraints on the EOS pressure (i.e., stiffness) at the highest NS densities, but this does not translate to any constraint on the TOV mass or any other macroscopic observable.
We do find an impact of the speed of sound at NS densities; it is widely accepted that the speed of sound (squared) increases beyond the conformal limit of 1/3 in densities relevant to NSs~\cite{Bedaque:2014sqa, Alsing:2017bbc, Tews:2018kmu, McLerran:2018hbz, Reed:2019ezm, Legred:2021hdx}, before decreasing to meet the 1/3 limit asymptotically (from below) at pQCD densities.
Under the pQCD constraints we find that this decrease starts occurring inside NSs.
The upper limit on the squared speed of sound at the center of the most massive NSs is $\lesssim 0.5$ at the 90\% credible level.
The impact of modeling assumptions on these results is explored in Appendix~\ref{app:gp2_flexibility}.
We conclude in Sec.~\ref{sec:discussion}.

\section{pQCD constraints}
\label{sec:pqcd_constraints}

We begin with a pedagogical overview of pQCD in Sec.~\ref{sec:pQCD_basics} and then describe how the relevant constraints can be incorporated into EOS inference in Secs.~\ref{sec:maximized_method} and~\ref{sec:modeled_method}.

\subsection{Perturbative QCD}
\label{sec:pQCD_basics}

Though quarks are usually confined to hadrons via the strong force, QCD predicts that at sufficiently high energy scales (i.e., sufficiently high densities) quark interactions become weaker---a phenomenon known as asymptotic freedom~\cite{Collins:1974ky}.
At such energy scales, a phase transition occurs to deconfined quark matter, corresponding to the ``color superconductor'' phase in the QCD phase diagram~\cite{Alford:2007xm, Fukushima:2010bq}.
In this regime, the strong coupling constant $\alpha_s$ (which determines the strength of the strong interaction) is small. 

Perturbative QCD amounts to an expansion in $\alpha_s$.
The leading-order term in the expansion for the pressure of the quark matter is that of a non-interacting Fermi gas governed by Fermi-Dirac statistics,
\begin{equation}
    p_\mathrm{FD} = \frac{3}{4\pi^2}\left(\frac{\mu}{3}\right)^4\,,
\end{equation}
where $\mu$ is the baryon chemical potential (equal to three times the quark chemical potential).
This expression directly leads to the expected behavior for the speed of sound at high densities, $\csq \rightarrow 1/3$.
Corrections to this leading-order behavior are due to interactions between the quarks.
Introducing a renormalization energy scale $\bar{\Lambda}$, a necessary ingredient in such calculations, we can expand the pressure in $\alpha_s(\bar{\Lambda})$ to give:
\begin{equation}
    p(\mu,\bar{\Lambda}) = p_\mathrm{FD}(\mu) + \sum_{n=1}^{\infty} \left(\alpha_s(\bar{\Lambda})\right)^n ~ p_n(\mu,\bar{\Lambda})\,.
    \label{eq:pqcd_expansion}
\end{equation}
The QCD strength $\alpha_s(\bar{\Lambda})$ is given by Eq.~(9) in Ref.~\cite{Kurkela:2009gj}, and depends on $\bar{\Lambda}$ such that $\alpha_s(\bar{\Lambda}) \rightarrow 0$ as $\bar{\Lambda} \rightarrow \infty$, i.e., as the energy or density goes to infinity.
Besides avoiding divergences in the truncated expansion, $\bar{\Lambda}$ also offers a convenient way to estimate the uncertainty in the (truncated) expansion from missing higher-order terms (in the limit of an infinite expansion, the dependence on $\bar{\Lambda}$ vanishes).
The expansion coefficients $p_n(\mu,\bar{\Lambda})$ for the (zero-temperature) pQCD EOS have been partially computed to $O(\alpha_s^3)$ (next-to-next-to-next-to-leading-order, N3LO) in Refs.~\cite{Gorda:2021znl, Gorda:2021kme, Gorda:2023mkk}.
The N3LO calculation is still ongoing, and builds on previous works~\cite{Freedman:1976ub, Baluni:1977ms, Kurkela:2009gj, Kurkela:2016was}; a summary of the terms in the expansion can be found in Ref.~\cite{Gorda:2023usm}.

\begin{figure}
    \centering
    \includegraphics[width=\columnwidth]{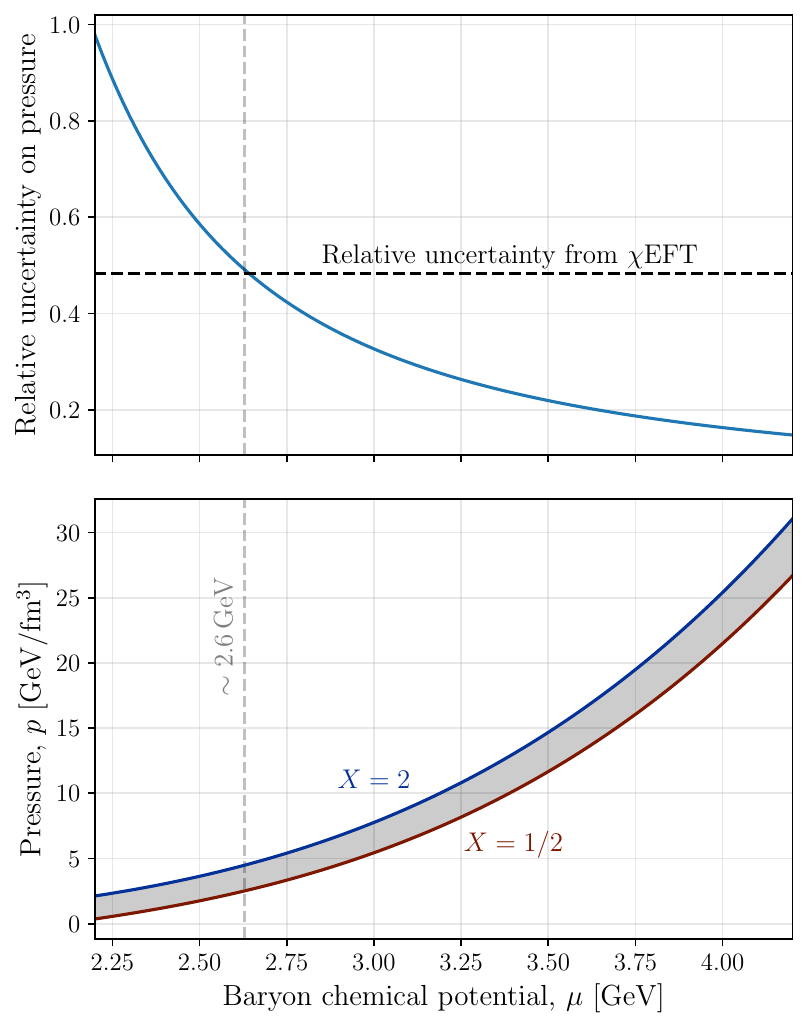}
    \caption{
    Pressure vs chemical potential from the pQCD expansion of Eq.~\eqref{eq:pqcd_expansion} (bottom panel), and the corresponding relative uncertainty on the pressure (top panel) when $X$ is varied between $1/2$ and 2.
    In the top panel we also indicate the relative uncertainty in the pressure from a $\chi$EFT calculation.
    As in Ref.~\cite{Komoltsev:2021jzg}, for the $\chi$EFT uncertainty, we take the difference in the ``soft'' and ``stiff'' EOSs at $1.1\,\nsat$ from Ref.~\cite{Hebeler:2013nza}.
    At around a chemical potential of $2.6\,\mathrm{GeV}$ the relative uncertainty from the pQCD calculation is equal to that of the $\chi$EFT calculation.
    }
    \label{fig:pqcd_uncertainty}
\end{figure}

The expansion becomes more accurate as $\bar{\Lambda} \rightarrow \infty$, which also corresponds to $\mu \rightarrow \infty$.
The renormalization scale is therefore closely related to (and often chosen to be proportional to) the chemical potential, since $\mu$ sets the energy scale of the system.
Following Ref.~\cite{Gorda:2023usm}, we adopt $\bar{\Lambda} = 2X\mu/3$ with $X \in [1/2, 2]$ controlling the uncertainty in the relation between $\bar{\Lambda}$ and $\mu$. 
This defines a region in the $\mu$--$p$ plane, as shown in the bottom panel of Fig.~\ref{fig:pqcd_uncertainty}, with $p(\mu, X=1/2)$ giving the lower bound on the pressure and $p(\mu, X=2)$ the upper bound.
This shaded region is the pQCD prediction for the EOS at high densities, akin to (for example) the $\chi$EFT envelope in Ref.~\cite{Hebeler:2013nza} for low densities.
The upper and lower bounds can be differentiated with respect to $\mu$ to give corresponding bounds on the number density, speed of sound, and other thermodynamic quantities.
For further details, see the discussion in Ref.~\cite{Gorda:2023usm}.

Given the pQCD prediction, \citet{Komoltsev:2021jzg} introduced the idea of using a ``pQCD likelihood'' to incorporate it into EOS inference. 
This likelihood can be applied to pre-existing EOS candidates (which themselves do not necessarily reach pQCD densities), and essentially asks how likely it is an EOS candidate can be connected to the pQCD prediction given some prescribed behavior at intermediate densities.
The original version of this likelihood has minimal assumptions about the EOS behavior, and we will refer to this as the ``maximized'' likelihood below. 
We describe this in Sec.~\ref{sec:maximized_method}.
The likelihood was then developed further by \citet{Komoltsev:2023zor} with a more constraining version which makes some assumptions about the EOS behavior, and involves conditioning a GP on the pQCD prediction. We will refer to this as the ``modeled'' likelihood, and it is described in Sec.~\ref{sec:modeled_method}.

\subsection{``Maximized'' pQCD constraint}
\label{sec:maximized_method}

\citet{Komoltsev:2021jzg} introduced the ``maximized'' pQCD constraint (code available at Ref.~\cite{komoltsev_2023_7781233}) that checks pre-existing EOS candidates (which themselves do not necessarily reach pQCD densities) against the pQCD prediction.
In practice, this check is often done at a single value for the chemical potential.
In the top panel of Fig.~\ref{fig:pqcd_uncertainty} we show the relative uncertainty on the pQCD pressure as a function of the chemical potential.
Based on this, a common choice is to check for pQCD consistency at a chemical potential of $\muH = 2.6\,\mathrm{GeV}$, for which the relative uncertainty on the pressure is comparable to the low-density $\chi$EFT uncertainty~\cite{Fraga:2013qra, Kurkela:2014vha}.
In the discussion of the method that follows we will also take a single value of the pressure $\pH$ for simplicity, but we note that uncertainty on $\pH$ can be accounted for by considering all values between $X = 1/2$ and $2$ (at fixed $\muH$).
We also have a number density $\nH$, related to the $\mu$--$p$ slope.
Higher derivatives, for example related to the speed of sound, do not play a role.

\begin{figure}
    \centering
    \includegraphics[width=\columnwidth]{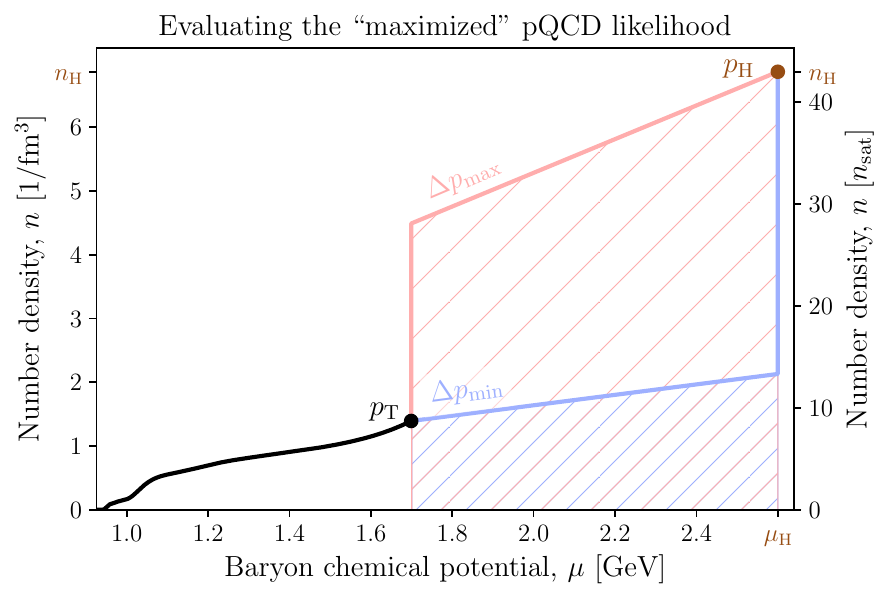}
    \caption{
    Given a pre-existing candidate EOS (black line) up to some termination point $p=\pT$ (black dot), the ``maximized'' pQCD constraint asks if it is possible to connect that EOS to the pQCD prediction at a single value $p=\pH$ (red dot). 
    This is done by considering EOSs that minimize (blue) and maximize (red) the pressure change from the termination point and checking if $\Delta p_\mathrm{min} \leq \pH - p_\mathrm{T} \leq \Delta p_\mathrm{max}$ (this is easily visualized in the $\mu$--$n$ plane, since the change in pressure corresponds to the area under the EOS).
    If this condition is true, then the EOS is given a likelihood of 1, else it is given a likelihood of 0.
    }
    \label{fig:constraint_evaluation}
\end{figure}

Given a pre-existing candidate EOS that terminates at $(\muT,\nT,\pT)$, the maximized constraint checks whether it is possible to connect that EOS to the pQCD prediction. 
To connect ($\muT,\nT,\pT$) and ($\muH,\nH,\pH$), we seek a continuous function of $\mu$ (thermodynamic consistency), and require that the number density $n$ is an increasing function of the chemical potential (thermodynamic stability). 
Additionally, the speed of sound
\begin{equation}
    \csq = \left(\frac{\mu}{n}\pdv{n}{\mu}\right)^{-1}\,,
\end{equation}
must be between 0 and 1, imposing a minimum gradient in the $\mu$--$n$ plane.
Thermodynamic consistency also implies that any function connecting these two points must accumulate the correct pressure,
\begin{equation}\label{eq:Delta p}
    \Delta p = \pH - \pT = \int_{\muT}^{\muH} \dd{\mu} n(\mu)\,.
\end{equation}
Given these considerations, we can construct two EOSs that link ($\muT,\nT$) and ($\muH,\nH$) while minimizing or maximizing the pressure change. 
How this is accomplished is shown in Fig.~\ref{fig:constraint_evaluation} with the blue (minimize the pressure change; $c_s=1$ at low $\mu$ followed by $c_s=0$ at high $\mu$) and red (maximize the pressure change; $c_s=0$ at low $\mu$ followed by $c_s=1$ at high $\mu$) lines for an example pre-existing EOS (black line).
When $c_s=0$, $\pdv*{\mu}{n} = 0$ and we can match any $n$ without changing $\mu$.
When $c_s=1$, $\mu \propto n$, and we can easily evaluate Eq.~\eqref{eq:Delta p}.
The minimum and maximum allowed pressure changes between ($\muT,\nT$) and ($\muH$,$\nH$) are therefore
\begin{equation}
    \Delta p_\mathrm{min} = \int_{\mu_\mathrm{T}}^{\muH} \dd{\mu} \frac{n_\mathrm{T}}{\mu_\mathrm{T}}\mu = \frac{1}{2}\frac{n_\mathrm{T}}{\mu_\mathrm{T}} \left( \muH^2 - \muT^2 \right)\,,
\end{equation}
and
\begin{equation}
    \Delta p_\mathrm{max} = \int_{\muT}^{\muH} \dd{\mu} \frac{\nH}{\muH}\mu = \frac{1}{2}\frac{\nH}{\muH} \left( \muH^2 - \muT^2 \right)\,,
\end{equation}
respectively. 
The pQCD ``maximized'' likelihood then performs the check
\begin{equation}\label{eq:maximized_likelihood}
    \Delta p_\mathrm{min} \leq \pH - p_\mathrm{T} \leq \Delta p_\mathrm{max}\,;
\end{equation}
if this condition is true, then it is possible to connect the pre-existing EOS to the pQCD prediction and it is assigned a likelihood of unity. 
If it is not true, the pre-existing EOS has a likelihood of zero.
Since this likelihood gives equal weight to all possible ways of connecting the pre-existing EOS to the pQCD prediction (no matter how extreme the behavior), it is the most conservative criteria for satisfying the pQCD constraints.
While flexible, this construction has a few drawbacks: (i) it considers the pQCD prediction at a single value of the chemical potential rather than the whole range of Fig.~\ref{fig:pqcd_uncertainty} (ii) it does not enforce a smooth transition for the speed of sound to the conformal limit, (iii) it weights all EOSs that can connect ($\muT,\nT,\pT$) and ($\muH,\nH,\pH$) equally, and (iv) it relies on a choice of where to terminate the existing EOS and start considering more extreme EOS behaviors.
Indeed, \citet{Komoltsev:2023zor} identified the latter as the source of the differences between Refs.~\cite{Gorda:2022jvk} and~\cite{Somasundaram:2022ztm}.

For a particular choice of the termination density $\nT$ and pQCD values $(\muH,\nH,\pH)$, Eq.~\eqref{eq:maximized_likelihood} defines an allowed region in the $\epsilon$--$p$ plane.
This is shown in the left panel of Fig.~\ref{fig:max_vs_marg}, where we have chosen $\nT = 10\,\nsat$, $\muH = 2.6\,\mathrm{GeV}$, and we have parametrized $\nH$ and $\pH$ via the dimensionless renormalization scale $X$ ($\epsilon_\mathrm{H}$ can be obtained via the relationship $\epsilon + p = \mu n$).
The region for $X=1/2$ is shown in red, and the region for $X=2$ is shown in blue.
If a candidate EOS is within this region at $10\,\nsat$ then it has likelihood unity, otherwise it has likelihood zero.
To avoid fixing $X$, it is common practice to marginalize over $X$ by sampling different values for $X$ from a log-uniform distribution between $1/2$ and 2, evaluating the likelihood for each, and then taking the average.
This is shown by the black shaded region, where darker regions indicate a higher likelihood. 
Appendix~\ref{app:epsilon-p} explains how to further add low-density $\chi$EFT constraints on the $p$--$\epsilon$ plane, thus elucidating Fig.~1 of Ref.~\cite{Komoltsev:2021jzg}, while Appendix~\ref{app:cs2} discusses how these results are altered when, instead of causality, we impose a different limit on the speed of sound.

\subsection{``Modeled'' pQCD likelihood}
\label{sec:modeled_method}

\begin{figure*}
    \centering
    \includegraphics[width=\linewidth]{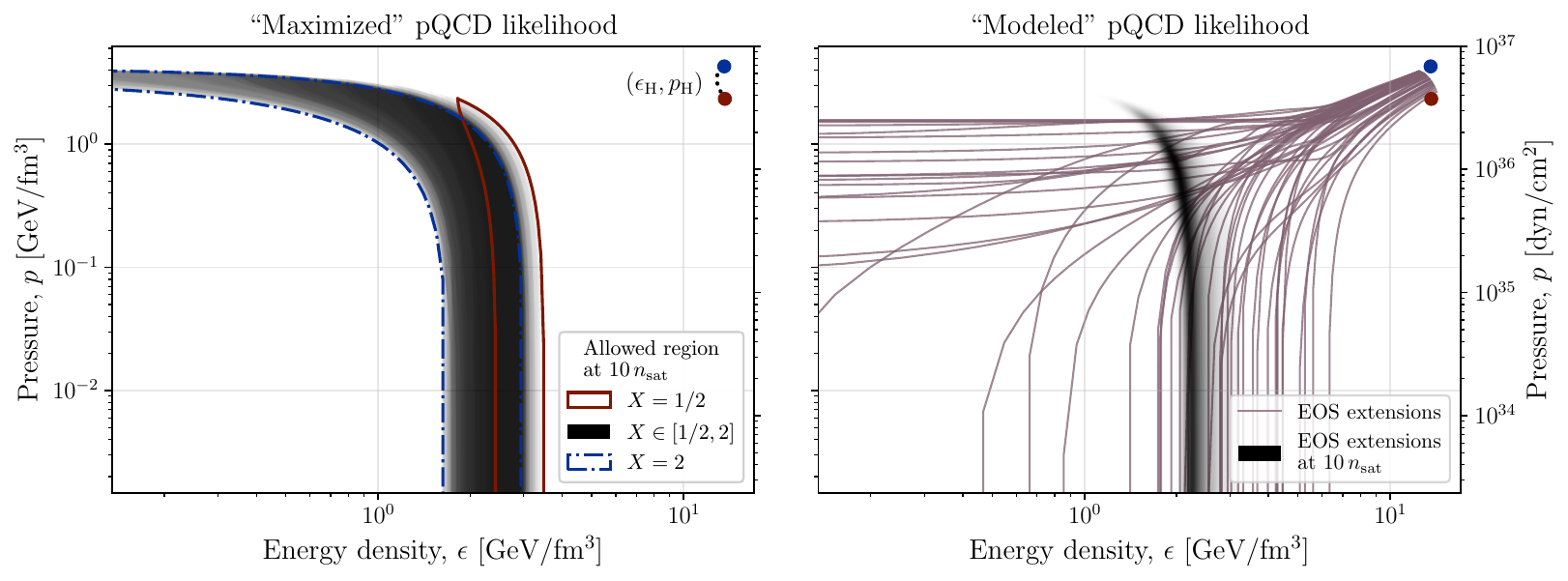}
    \caption{
    Visualization of the ``maximized'' (left) and ``modeled'' (right) pQCD likelihoods with $\nT = 10\,\nsat$.
    (Left panel) For a particular choice of the renormalization scale $X$ and $\muH$ (dots), the maximized pQCD likelihood, Eq.~\eqref{eq:maximized_likelihood}, defines an allowed region in the $\epsilon$--$p$ plane for the EOS at $\nT$.
    We show this region for $X = 1/2$ (red) and $X = 2$ (blue), both with $\muH = 2.6\,\mathrm{GeV}$.
    Marginalizing over X with a log-uniform distribution between $1/2$ and $2$, we obtain the black shaded region with darker regions having a higher likelihood. 
    (Right panel) The modeled likelihood instead uses EOS extensions that start at the pQCD high-density prediction (whilst taking into account uncertainty on $X$) and are propagated to lower densities.
    A sample of these EOSs are shown in pink, taken from the ``conditioned'' GP of Ref.~\cite{Komoltsev:2023zor} (which also imposes $\csq \rightarrow 1/3$ at densities above $25\,\nsat$).
    The point where these EOS extensions reach $\nT$ is identified, and a KDE is built on the corresponding pressure and energy-density values, shown with the shaded black region.
    }
    \label{fig:max_vs_marg}
\end{figure*}

\citet{Komoltsev:2023zor} proposed an alternative method that introduces an explicit model for the EOS between the termination point of the pre-existing EOS candidate and the pQCD prediction (code available at Ref.~\cite{komoltsev_2025_15407795}).
The EOS model is based on a GP that is conditioned on the whole pQCD prediction in $n=(25,40)\,\nsat$, roughly corresponding to $\mu>2.2\,$GeV, thus also reproducing the expected $\csq \rightarrow 1/3$ limit.
A sample of these GP EOSs are shown in the right panel of Fig.~\ref{fig:max_vs_marg}, starting from the high-density pQCD prediction and propagating to lower densities.

Given, again, a pre-existing low-density candidate EOS that terminates at $(\muT,\nT,\pT)$, the modeled likelihood checks consistency against the high-density GP EOSs that already satisfy pQCD.
The procedure is the following.
The energy density and pressure of the GP EOS extensions at $\nT$ are identified and a kernel density estimate (KDE) is built on these values.
This KDE is the new ``modeled'' pQCD likelihood, and is shown with the black shaded region in the right panel of Fig.~\ref{fig:max_vs_marg} for the choice $\nT = 10\,\nsat$.
The likelihood naturally disfavors ``extreme'' EOSs with fine-tuned behavior (for example, EOSs with strong phase transitions or extended regions of $c_s = 1$), since these constitute a small fraction of the parameter space that the GP explores.
Compared with the maximized likelihood on the left panel of Fig.~\ref{fig:max_vs_marg}, the modeled likelihood has a reduced range of support; a result of down-weighting extreme EOSs and also conditioning on the full pQCD prediction.

While more physically informed than the ``maximized'' constraint, the ``modeled'' constraint still has some drawbacks.
Firstly, it does not do away with the need to select a termination point $\nT$ at which to apply the likelihood.
Secondly, with an explicit model for the post-$\nT$ EOS, the result will now depend on assumptions for the GP EOS extension behavior.
For example, the EOS extensions of Ref.~\cite{Komoltsev:2023zor} rely on square-exponential kernels with length scale and variance (drawn uniformly and normally, respectively) as $\ell \sim \mathcal{U}(1\,\nsat,20\,\nsat)$, $\eta \sim \mathcal{N}(1.25, 0.25^2)$,
and uses a constant mean (itself drawn from a normal distribution).
These values result in relatively smooth EOSs, cf., the right panel of Fig.~\ref{fig:max_vs_marg}.
Thirdly, related to the previous points and discussed in \citet{Komoltsev:2023zor} (see also Sec.~2.4c of \citet{Komoltsev:2025vwn}), the modeled likelihood introduces a second-order phase transition (drop in sound speed) when applied at $\nTOV$, which is not necessarily desired behavior.
And finally, both the maximized and modeled approaches suffer from the fact that $\MTOV$ may change based on the extension (due to the presence of exotic stable branches~\cite{Essick:2024olf}), and therefore it may not be self-consistent to use this extension alongside constraints based on TOV quantities derived from only the low-density EOS.
In Sec.~\ref{sec:pqcd_prior} we introduce a unified EOS that spans the entire relevant density range, directly addressing the first, third, and fourth points above.
Like the EOS extensions of the modeled pQCD likelihood, the unified EOS is based on a GP and so will inevitably depend on assumptions made for the GP EOS behavior.
However, care is taken to ensure our EOS is sufficiently flexible to capture a wide range of EOS behaviors (and a wider range of behaviors than the EOS extensions).

\begin{figure*}
    \centering
    \includegraphics[width=\linewidth]{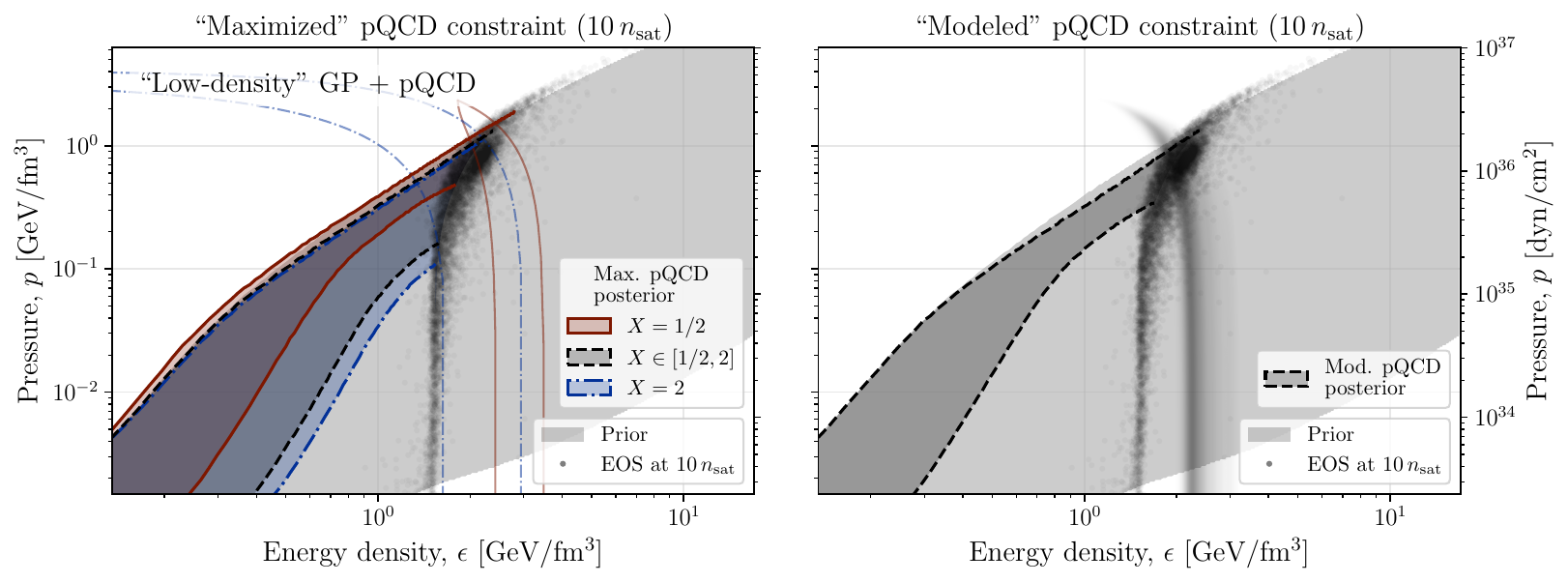}
    \caption{
    Posteriors for the EOS pressure and energy density (90\% credible intervals) from applying the maximized (left) and modeled (right) pQCD constraints at $\nT=10\,\nsat$.
    Posteriors are truncated at that value, beyond which the EOS model changes. 
    The energy density and pressure of individual EOSs at $10\,\nsat$ are highlighted with black markers.
    The prior is shown in light gray; it contains no astrophysical input.
    These results demonstrate which EOSs are being ruled out by pQCD alone.
    (Left panel) As in Fig.~\ref{fig:max_vs_marg}, we consider the maximized constraint at $X=1/2$ (red), 2 (blue), and marginalized over $X$ (black dashed). 
    We also show the $X=1/2$ and $X=2$ allowed regions for the maximized likelihood from Fig.~\ref{fig:max_vs_marg} (thin red and blue lines). 
    To satisfy the maximized pQCD likelihood, the EOS at $10\,\nsat$ must be within the allowed region.
    (Right panel) As in Fig.~\ref{fig:max_vs_marg}, we include the modeled pQCD likelihood represented as a KDE.
    Only EOSs which overlap with this KDE at $10\,\nsat$ have non-zero weight.
    }
    \label{fig:pqcd_posterior}
\end{figure*}

\section{The impact of pQCD constraints}
\label{sec:post-hoc}

In this section we explore the impact of existing formulations of the pQCD constraints on dense matter.
Following Refs.~\cite{Gorda:2022jvk, Somasundaram:2022ztm, Komoltsev:2023zor, Koehn:2024set} we consider the maximized and modeled constraints applied at some termination density $\nT$.
This can also be viewed as a transition density, since above $\nT$ we are switching to a different EOS model assumed in the construction of the pQCD likelihood.
Below $\nT$, the EOS is described by realizations from the GP constructed in Refs.~\cite{Landry:2018prl, Essick:2019ldf, Landry:2020vaw}, referred to as the ``low-density'' GP, to distinguish it from the ``unified'' GP in the subsequent section.
The GP is constructed with a focus on flexibility and probing a wide range of possible behavior, including phase transitions~\cite{Essick:2023fso}.
EOSs are smoothly attached to a model for the crust at low density~\cite{Baym:1971pw, Hebeler:2013nza}, and then designed to explore all causal and thermodynamically stable behaviors at high density~\cite{Legred:2022pyp}.
This is achieved via a mixture of kernels exploring a range of correlations with length scale and variance chosen using a cross-validation likelihood
~\cite{Essick:2019ldf} (see also Appendix A of \citet{Legred:2022pyp}).
Consequently, draws from the low-density GP are not consistent with the pQCD constraints \textit{a priori}, which are instead applied at some choice of $\nT$.
In Sec.~\ref{sec:pqcd_posterior} we investigate the impact of the pQCD constraints when applied on their own to the very flexible low-density GP prior.
In Sec.~\ref{sec:synergy} we consider pQCD in conjunction with astrophysical information.

\begin{figure*}
    \centering
    \includegraphics[width=\linewidth]{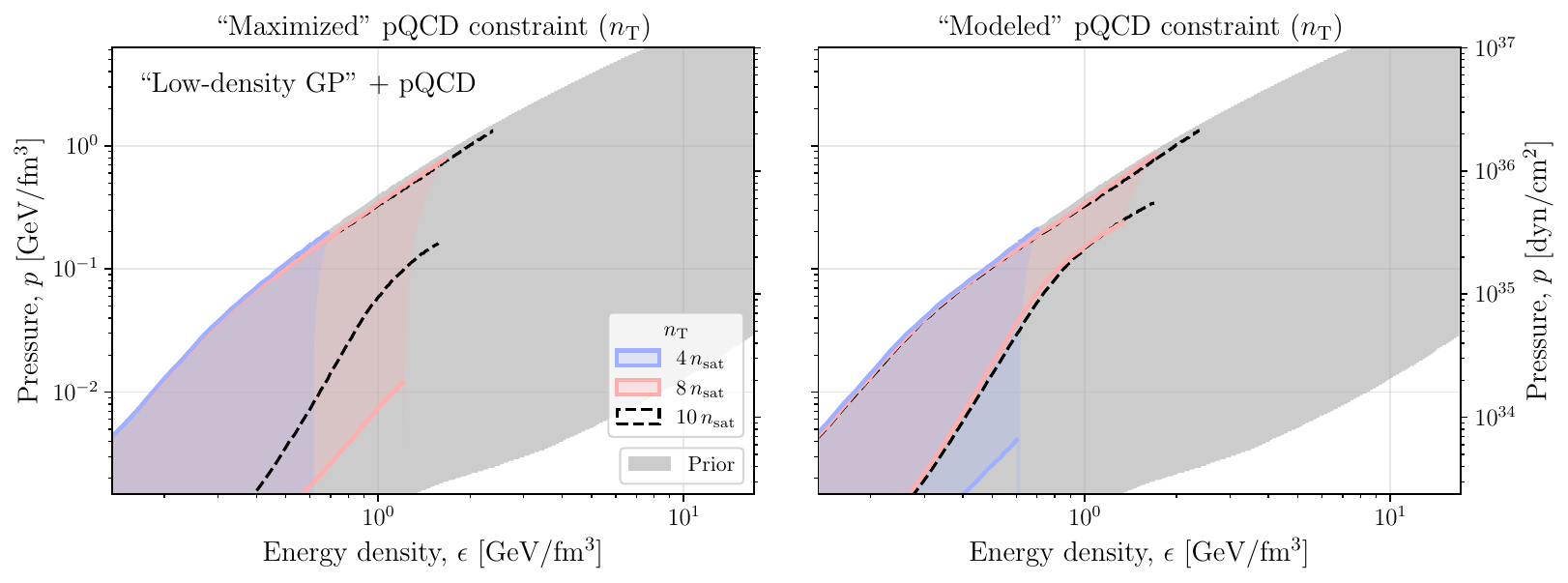}
    \caption{
    Similar to Fig.~\ref{fig:pqcd_posterior}, but for other choices of the termination density $\nT$.
    We show the prior (gray shaded region) and posteriors for different values of $\nT$ (pink, blue, and black) for the maximized (left) and modeled (right) constraint.
    For the maximized constraint we have marginalized over $X$.
    This comparison quantifies the impact of $\nT$ on the posterior, i.e., the value of the density at which we switch from the low-density GP EOS model to the high-density EOS that is tailored to the pQCD limit.
    \label{fig:pqcd_posterior_nterm}
    }
\end{figure*}

\subsection{pQCD posterior from a broad prior}
\label{sec:pqcd_posterior}

To study the raw impact of pQCD constraints on dense matter, we first apply them on their own to the broad low-density GP prior, without any input from astronomical observations. 
Figure~\ref{fig:pqcd_posterior} shows results for the pressure as a function of energy density for $\nT=10\,\nsat$. 
The posteriors are a combination of the likelihoods shown in Fig.~\ref{fig:max_vs_marg} (consistent color scheme) and the low-density GP prior shown in faint gray.
All distributions here and throughout are plotted at the 90\% credible level.
For the ``maximized'' constraint (left panel) we show the posterior for fixed $X = 1/2$ and $2$ (red and blue shaded regions, respectively), as well as marginalized over $X$ (region marked by dashed black lines). 
Black markers indicate the values of individual EOSs at $10\,\nsat$. 
For fixed $X$, any EOS whose marker does not fall within the allowed region (thin red and blue lines, same as in Fig.~\ref{fig:max_vs_marg}) is rejected.
Rejected EOSs fall into two groups: the ones with the highest pressure and energy density at $10\,\nsat$ and the ones with the lowest pressures.
It is the high pressure/energy density EOSs (i.e., those that fall to the right of the allowed region) that have been previously shown to be ruled out, leading to the softening of the EOS~\cite{Gorda:2022jvk}.
As will be explored later, the low pressure EOSs that are ruled out are also largely ruled out by the astrophysical data.
The EOS behavior should not be considered above where the pQCD likelihood is applied, $\nT$, and so we truncate the posteriors at $\nT$.
This can also be viewed as the density where the EOS model switches from the low-density GP to the maximally-agnostic construction of Sec.~\ref{sec:maximized_method}.
Consistent with previous work~\cite{Somasundaram:2022ztm}, constraints are strongest for lower values of $X$, which is a consequence of the pQCD pressure and number density being lower for lower values of $X$ (meaning there is less parameter space available to be consistent with the constraints).

The right panel of Fig.~\ref{fig:pqcd_posterior} shows the results after applying the ``modeled'' constraint. 
We again show the prior, posterior truncated at $\nT$, markers denoting EOS values at $10\,\nsat$, and additionally include the same KDE representing the likelihood as in Fig.~\ref{fig:max_vs_marg}. 
For each EOS, the pQCD constraint now amounts to the probability of drawing its associated marker value from the likelihood KDE. 
The modeled constraint provides a qualitatively similar but tighter bound than the maximized one since it down-weights EOSs with extreme behavior such as strong phase transitions and a causal sound speed.
The modeled constraint also incorporates the requirement that the pQCD speed of sound starts to tend toward $\csq \rightarrow 1/3$ from densities of $\sim 25\,\nsat$, which further limits the allowed EOS behavior.

It is widely appreciated that these results depend on the choice of $\nT$~\cite{Komoltsev:2021jzg,Komoltsev:2023zor}. 
Figure~\ref{fig:pqcd_posterior_nterm} explores this dependence for the pressure vs energy-density posterior.
We plot the prior and the posterior at different values of $\nT$, with each posterior again terminating at that value.
For both the maximized (left panel) and modeled (right panel) constraints, decreasing the value of $\nT$ severely weakens the constraints. 
This is especially true for the more agnostic maximized constraint, where by $\nT=4\,\nsat$ the posterior is almost identical to the prior.
This is consistent with previous works~\cite{Komoltsev:2021jzg,Komoltsev:2023zor}, and is simply a consequence of a wider region of the phase space being consistent with pQCD at lower densities.

The quantitative dependence on the value of $\nT$ introduces a level of arbitrariness in both the maximized and modeled constraints.
Since the EOS model effectively changes at $\nT$ it is not useful to apply either constraint at or below densities relevant for NSs; placing the switching of EOS representations in a region relevant to NS interiors means that the resulting constraints cannot be used to explain astronomical NSs.
The choice $\nT=10\,\nsat$ avoids this problem as it corresponds to the upper tail of the $\nTOV$ distribution for the low-density GP~\cite{Legred:2021hdx}, but this choice may be overly restrictive since it means applying the constraint a few $\nsat$ above $\nTOV$ for the majority of EOS.
Motivated by this, another choice is $\nT=\nTOV$~\cite{Somasundaram:2022ztm, Koehn:2024set,Mroczek:2023zxo} (so that the constraint is applied at a different density for each EOS candidate).
This is the most conservative choice possible whilst still being able to study densities relevant for NSs, and is the choice we make in the following sections.
However, as mentioned earlier, this choice still allows for an abrupt change in EOS behavior immediately after $\nTOV$.
Since this is the maximum density of the most massive \textit{nonspinning} NS~\cite{Komoltsev:2023zor}, it thus does not consider spinning~\cite{Paschalidis:2016vmz} or hypermassive~\cite{Baumgarte:1999cq, Kastaun:2014fna, Hanauske:2016gia} stars that can reach a higher mass.

\begin{figure*}
    \centering
    \includegraphics[width=\linewidth]{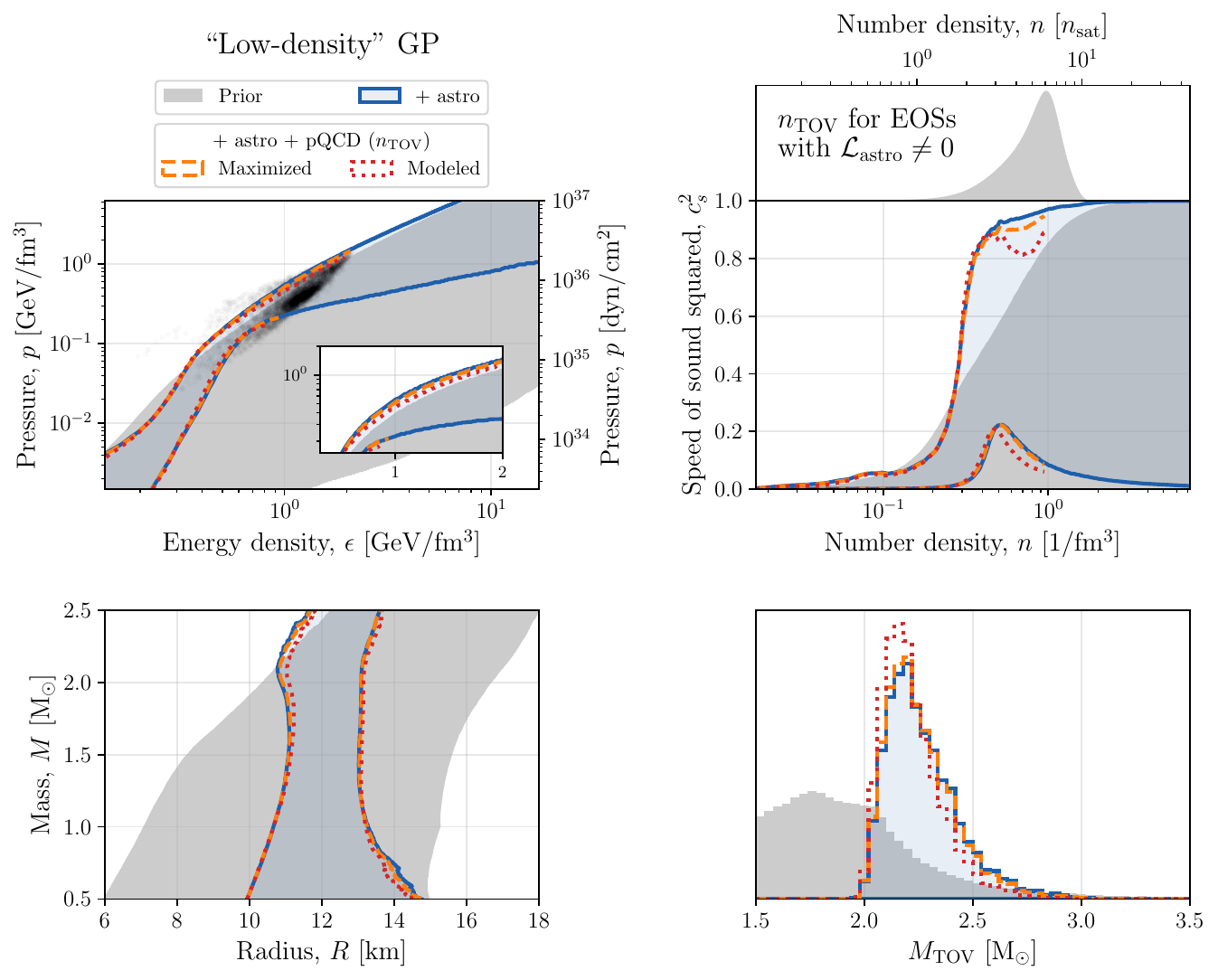}
    \caption{
    EOS prior (gray shade) and posteriors (90\% credible intervals) from application of the astrophysical data (blue solid) and additionally pQCD maximized (orange dashed; marginalized over $X$) and modeled (red dotted) constraints at $\nT=\nTOV$.
    We show the pressure as a function of the energy density (top left), the speed of sound as a function of the number density (top right), the NS mass and radius (bottom left), and $\MTOV$ (bottom right).
    In the top left panel we indicate with black markers the pressure and energy density for a sample of EOSs at $\nTOV$ (beyond which the posteriors with pQCD applied should not be considered) and in the top right panel we show the distribution on $\nTOV$; for both we only use EOSs with non-zero astrophysical likelihood to remove unphysical EOSs.
    }
    \label{fig:astro_pqcd_posterior_gp0}
\end{figure*}

\subsection{Synergy with astrophysical constraints}
\label{sec:synergy}

At densities $1 \text{--} 10\,\nsat$ that are most relevant to NSs, most information on the EOS comes from astronomical observations. 
It is thus expected that a portion of the parameter space ruled out by pQCD in Fig.~\ref{fig:pqcd_posterior}, and especially the lowest pressures, has already been ruled out by other observations~\cite{Somasundaram:2022ztm}.
We explore the relation between astrophysical and pQCD constraints in this section, incorporating the former through the framework laid out in Refs.~\cite{Landry:2020vaw,Chatziioannou:2020pqz}. 
Following Refs.~\cite{Landry:2020vaw,Legred:2021hdx,Golomb:2024lds}, we consider radio, X-ray, and gravitational-wave data.
Radio data consist of a mass measurement for J0348+0432~\cite{Antoniadis:2013pzd} and J0740+6620~\cite{NANOGrav:2019jur,Fonseca:2021wxt} which impose a lower limit on $\MTOV$.
X-ray data consist of mass and radius measurement for J0030+0451~\cite{Miller:2019cac, Riley:2019yda}, J0740+6620~\cite{Miller:2021qha, Riley:2021pdl} (combined with radio data), and J0437–4715~\cite{Choudhury:2024xbk}.
For J0030+0451 and J0740+6620 we use results from Refs.~\cite{Miller:2019cac,Miller:2021qha}, and for J0030+0451 we use the model with three oval spots.
Gravitational-wave data consist of mass and tidal deformability measurements for the two binary components of GW170817~\cite{LIGOScientific:2017vwq} and GW190425~\cite{LIGOScientific:2020aai}.
For GW170817 we use the results corresponding to the ``low-spin'' prior of Ref.~\cite{LIGOScientific:2018hze} (data available at Ref.~\cite{gw170817_data}), and for GW190425 we use the ``high-spin'' prior (data available at Ref.~\cite{gw190425_data}).

\begin{figure*}
    \centering
    \includegraphics[width=\linewidth]{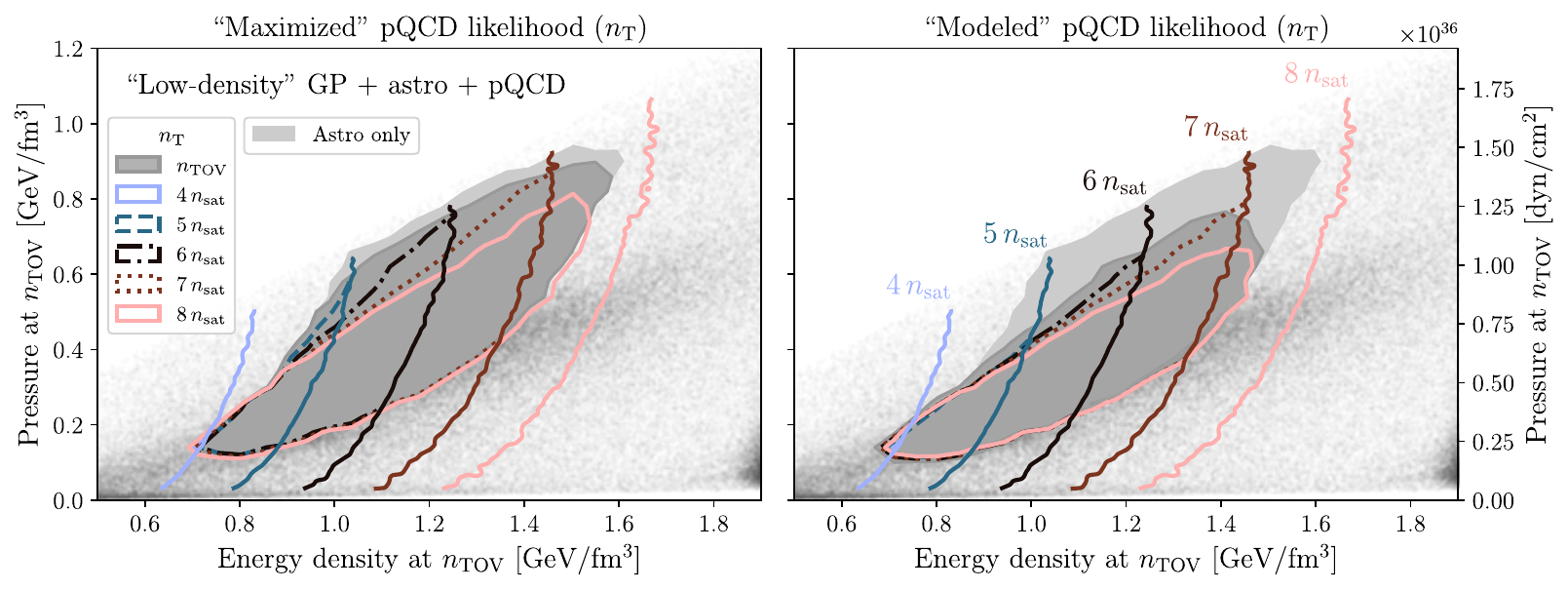}
    \caption{
    Posteriors (90\% credible regions) for the pressure and energy density at $\nTOV$ when astronomical constraints are applied alongside the maximized (left, marginalized over $X$) and modeled (right) pQCD constraints for a selection of termination densities $\nT$.
    We show fixed values in increments of $\nsat$ (different colors, empty contours) as well as $\nT=\nTOV$ (dark gray filled contour).
    The light gray filled contour corresponds to the posterior without any pQCD input.
    Background dots indicate draws from the prior.
    Each EOS reaches $\nTOV$ at a different density, but $\nTOV$ generally increases with energy density; the labeled upward lines indicate the locus of constant $\nTOV$.
    Since the EOS posterior from low-density GP should not be considered above $\nT$, the part of the contour that is to the right of the corresponding line should not be considered and thus we truncate the contours at $\nT$.
    }
    \label{fig:pqcd_astro_e_p_tov_posterior}
\end{figure*}

Figure~\ref{fig:astro_pqcd_posterior_gp0} shows results when the maximized and modeled pQCD constraints are applied at $\nT=\nTOV$. 
We show the prior, posterior using only astronomical data, and posteriors when additionally including pQCD for the pressure as a function of the energy density (top left), the speed of sound as a function of the number density (top right), the NS mass and radius (bottom left), and $\MTOV$ (bottom right).
The astro-only posterior (blue solid) is similar to results presented in numerous studies with the low-density GP~\cite{Landry:2020vaw, Legred:2021hdx} with $R_{1.4}=12.07^{+0.96}_{-1.03}\,$km, $R_{2.0}=12.07^{+1.02}_{-1.20}\,$km, and $\MTOV = 2.24^{+0.36}_{-0.17}\,\mathrm{M}_\odot$ at the 90\% credible level.
Switching to the impact of pQCD, the maximized pQCD constraint when applied at $\nT=\nTOV$ offers minimal additional information on top of current astrophysical data in all panels (orange dashed), and for both microscopic (top) and macroscopic (bottom) quantities.
This result is consistent with Refs.~\cite{Somasundaram:2022ztm},~\cite{Mroczek:2023zxo}, and~\cite{Koehn:2024set}, who primarily employed the maximized pQCD likelihood and saw minimal impact (see, for example, the top row of \citet{Koehn:2024set} Fig.~28).

As before, the modeled constraint is slightly stronger, and looking at the top left panel of Fig.~\ref{fig:astro_pqcd_posterior_gp0} we see the upper limit on the pressure at $\epsilon= 1\,\mathrm{GeV/fm}^3$ reduces by $\sim 20\%$ compared to the astro-only constraint.
This result corresponds to the general ``softening'' effect of pQCD on the high-density EOS, e.g., Ref.~\cite{Gorda:2022jvk}.
This is clearer in the right panel of Fig.~\ref{fig:pqcd_astro_e_p_tov_posterior}, where we plot the posterior on the pressure and energy density at $\nTOV$.

Turning to the speed-of-sound posterior (Fig.~\ref{fig:astro_pqcd_posterior_gp0}, top right panel), we see that pQCD information rules out the highest values for the speed of sound at densities close to $\nTOV$ (we show a KDE of $\nTOV$ values for reference). 
Specifically, with only astrophysical constraints, the squared speed of sound at $\nTOV$ is $c_{s,\mathrm{TOV}}^2 = 0.52^{+0.44}_{-0.42}$, whilst with the ``modeled'' pQCD constraints applied at $\nT=\nTOV$ we have $c_{s,\mathrm{TOV}}^2 = 0.35^{+0.52}_{-0.28}$.
For the maximum squared speed of sound reached anywhere inside a NS, for the modeled constraint we see a slight decrease from $c_{s,\mathrm{max}}^2 = 0.85^{+0.14}_{-0.31}$ (with only astrophysical constraints) to $c_{s,\mathrm{max}}^2 = 0.78^{+0.22}_{-0.27}$ and it is pushed from a density of $3.71^{+2.96}_{-1.71}\,\nsat$ to $3.07^{+3.14}_{-1.17}\,\nsat$. 
At a density of $\sim 5\,\nsat$, we see the upper bound of the speed of sound (with the modeled pQCD constraint applied) starts to increase again.
This feature can be seen in other works; for example, see the left panel of Fig.~5 from \citet{Gorda:2022jvk} and Fig.~4 from \citet{Somasundaram:2022ztm}.
Both of these works applied the ``maximized'' pQCD constraint, and we have confirmed we see the same behavior with the maximized constraint applied at fixed termination densities.
This is likely due to the fact that the pQCD constraints are applied in the pressure vs energy density space, and so it is the derivative of the constraints that apply in the speed of sound vs number density space (i.e., it is an integral of the speed of sound that is being constrained); this means that the speed of sound just below $\nT$ is only weakly influenced by the pQCD constraints.

The top two panels of Fig.~\ref{fig:astro_pqcd_posterior_gp0} suggest that pQCD mostly informs the highest NS densities, especially after astronomical data are considered.
The bottom panels confirm this picture from the macroscopic point of view.
The NS radius is largely unaffected for masses $\lesssim 1.6\,\mathrm{M}_{\odot}$, but the radius of a $2\,\mathrm{M}_{\odot}$ star is constrained from below, with the 90\% lower limit showing a small increase from $10.9\,\mathrm{km}$ to $11.1\,\mathrm{km}$ when the modeled constraint is applied.
We find a mild reduction in $\MTOV$ under the modeled constraint, dropping from $2.24^{+0.36}_{-0.17}\,\mathrm{M}_\odot$ to $2.19^{+0.29}_{-0.15}\,\mathrm{M}_\odot$, again consistent with Ref.~\cite{Koehn:2024set}---see their Appendix D.
The exact posterior on $\MTOV$ will depend on the prior, and this should be taken into account when comparing with other works.
For example, \citet{Gorda:2022jvk} have a prior on $\MTOV$ which is roughly uniform, whilst ours is peaked at two solar masses (our prior still has support up to three solar masses, and so it is sufficiently wide to resolve posterior differences).

The increase in $R_{2.0}$ under the modeled pQCD constraint appears at first inconsistent with the fact that the EOS is ``softened,'' and thus the pressure decreases, at high densities.
This is because radii are positively correlated with the pressure at twice saturation~\cite{Lattimer:2000nx,Legred:2023als}, and, in fact, we see a modest decrease in the radius at low masses when pQCD constraints are applied.
However, this trend reverses with the higher, core pressures when conditioning on astrophysical data.
For example, scaling relations for polytropic EOSs show that gravitating objects with larger radii will have lower central pressures~\cite{Shapiro:1983du, Silbar:2003wm}. 
EOSs that can produce high-mass stars will therefore tend to have larger radii.
As such, the pQCD constraints, which place upper bounds on the central pressure and energy density (see Fig.~\ref{fig:pqcd_astro_e_p_tov_posterior}), effectively impose a lower bound on the radii of high-mass NSs.

The EOS properties close to the TOV limit are further explored in Fig.~\ref{fig:pqcd_astro_e_p_tov_posterior} with the $p_{\mathrm{TOV}}$--$\epsilon_{\mathrm{TOV}}$ posterior.
We show the prior with background dots, the posterior when only including astrophysical data, and the posterior when further adding pQCD constraints at different values of $\nT$. 
Upward lines show the locus of constant $\nTOV$, marking the highest density at which the corresponding posterior contour is applicable.
For example, the black dash-dotted contour for $\nT=6\,\nsat$ is only applicable to the left of the black line marking the locus of $6\,\nsat$.
To aid interpretation, we have truncated the contours at the corresponding line.
For $\nT=\nTOV$ and $\nT = 8\,\nsat$, the entire posterior contour is applicable.

For the maximized constraint (left), we again confirm that the chosen value of $\nT$ impacts the posterior, with higher values leading to stronger constraints in terms of limiting the pressure, e.g., Refs.~\cite{Komoltsev:2023zor, Somasundaram:2022ztm, Mroczek:2023zxo} and Fig.~\ref{fig:pqcd_posterior_nterm}.
Moreover, the maximized pQCD constraint applied at $\nT=\nTOV$ is minimally informative on top of astrophysical data.
The picture is different for the modeled constraint (right) which is slightly stronger.
As already shown in Fig.~\ref{fig:astro_pqcd_posterior_gp0}, for energy densities close to the TOV limit, pQCD rules out the highest pressures compared to astrophysical data.
The central pressure of the maximum-mass TOV star reduces from $p_\mathrm{TOV}=0.49^{+0.36}_{-0.28}\,\mathrm{GeV/fm}^3$ when only astronomical data are used to $p_\mathrm{TOV}=0.36^{+0.33}_{-0.19}\,\mathrm{GeV/fm}^3$ with pQCD information.
And as already shown in Fig.~\ref{fig:pqcd_posterior_nterm}, the modeled constraint is less sensitive to the choice of $\nT$ (particularly when used in conjunction with astrophysical information).
The truncated contours visualize and elucidate this result.
At a given $\nTOV$ (indicated by the roughly vertical lines), applying the maximized constraint (left) at higher $\nT$ results in a narrower posterior for $p(n_\mathrm{TOV})$, whereas for the modeled constraint (right) the posterior credible intervals largely overlap.
This is because the effective EOS model post-$\nT$ is much smoother than the pre-$\nT$ EOS model between $4$--$6\,\nsat$ for the modeled case than the maximized case.
Since the total $\nT=\nTOV$ constraint is a convolution of the constraints at multiples of $\nsat$ with the EOS central density, the $\nT=\nTOV$ posterior ends up consistent with the posteriors at a fixed value of $\nT$, for densities up to that value.

\section{A unified EOS model up to pQCD densities}
\label{sec:pqcd_prior}

Both the maximized and modeled pQCD constraints explored in Sec.~\ref{sec:post-hoc} require a somewhat arbitrary choice of the termination density $\nT$ and depend on the allowed EOS behavior post-$\nT$, which might introduce inconsistencies in determining $M_\mathrm{TOV}$~\cite{Essick:2024olf}.
In this section, we introduce a ``unified'' EOS model that extends from the NS crust to densities applicable to pQCD.
While the idea of coherently modeling the EOS from the NS crust to densities relevant for pQCD is not new~\cite{Kurkela:2014vha, Fraga:2015xha, Annala:2017llu, Annala:2019puf, Annala:2021gom, Altiparmak:2022bke, Shirke:2022tta, Semposki:2024vnp, Semposki:2025etb}, our approach makes use of a unique GP at intermediate densities to maximize model flexibility.
As already discussed in the context of the modeled constraints, the use of a GP is not a panacea: the details of the GP and its level of flexibility still affect the results.
Appendix~\ref{app:gp2_flexibility} discusses the flexibility of the unified GP.
This approach, however, allows us to dispense with the need to choose an $\nT$ and change the EOS behavior across it.
Along with removing an arbitrary choice, this is also advantageous in terms of interpretability; since we have a single model across all densities, any impact at NS densities can be understood in terms of the EOS behavior at higher densities.

\begin{figure}[b]
    \centering
    \includegraphics[width=\columnwidth]{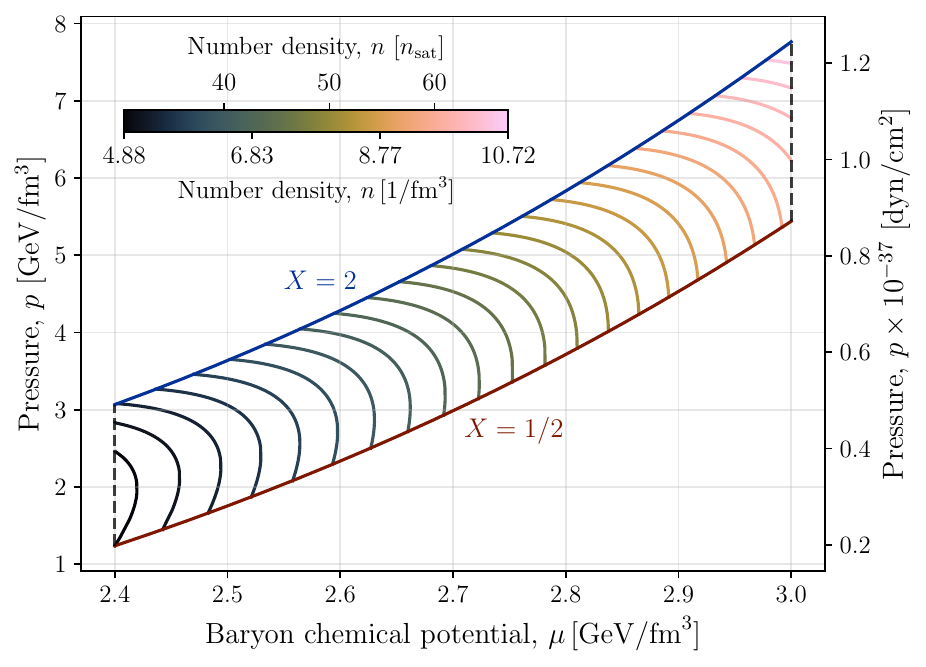}
    \caption{
    Visualization of the ``pQCD window'' that we require the EOS draws to hit. 
    When the renormalization scale $X$ is varied alongside the pQCD chemical potential, this defines a surface in the pressure vs number density vs chemical potential space.
    }
    \label{fig:pqcd_window}
\end{figure}

\subsection{Unified GP construction}
\label{sec:unified_construction}

Our goal is to explicitly model the EOS up to pQCD densities and ensure that it reaches pQCD-consistent values.
This can be thought of as incorporating the pQCD information into the prior, much like how at low density GPs can be conditioned on $\chi$EFT~\cite{Essick:2020flb,Gorda:2022jvk} or relativistic mean-field theory~\cite{Legred:2025aar}.
In practice, however, it is nontrivial to construct a GP that simultaneously satisfies given low-density (corresponding to a crust model) and high-density (corresponding to pQCD) behavior for the following reason.

Say we seek a GP from which draws reach $(\muL,\nL,\pL)$ at low densities and $(\muH,\nH,\pH)$ at high densities whilst satisfying causality and thermodynamic stability. 
If we were to construct the GP in the $\mu$--$n$ space, we could easily obtain EOSs that pass through $(\muL,\nL)$ and $(\muH,\nH)$ via conditioning.
However, we would still need to impose a minimum gradient (causality) and a fixed area under the line (pressure change from $\pL$ to $\pH$)---these are hard to satisfy simultaneously.
Instead, Refs.~\cite{Landry:2018prl, Essick:2019ldf, Landry:2020vaw} prioritize causality by working in the $\phi$--$p$ space, where $\phi$ is an auxiliary variable (related to the speed of sound) that maps between the GP domain $\phi \in [-\infty,\infty]$ and the causal range $\csq \in [0,1]$ (this can also be done in the $\phi$--$n$ space~\cite{Gorda:2022jvk, Komoltsev:2023zor, Koehn:2024set}).
Via $\phi$ we can control the speed of sound at any pressure via conditioning, meaning we can match low- and high-density speed-of-sound behavior simultaneously.
Converting $\phi$--$p$ draws to chemical potential, energy density, and pressure requires integration, e.g., Eqs.~(8) and (9) in Ref.~\cite{Gorda:2022jvk}, which introduces two integration constants.
We can choose these constants to match either the low- or high-density behavior, but not both.
The ``low-density'' GP of Sec.~\ref{sec:post-hoc} fixes the low-density crust behavior, while the GP of Ref.~\cite{Komoltsev:2023zor} from which the modeled constraint originates fixes the high-density behavior and thus automatically satisfies pQCD. 
In other words, it is not possible within current GP parametrizations to create draws that automatically satisfy both criteria simultaneously.
Alternative parametrizations might resolve this issue.
Lacking such an alternative, we instead continue building the GP in the $\phi$ (speed of sound) vs pressure space~\cite{Landry:2018prl, Essick:2019ldf, Landry:2020vaw}, fix the low-density limit through the integration constants, and explicitly check each draw for pQCD consistency.

In order to determine if a given EOS is consistent with pQCD, it is useful to think of the EOS as a line in the three-dimensional space of pressure, number density, and chemical potential.
At a given chemical potential, varying the renormalization scale $X$ in the pQCD calculation defines a line in the pressure vs number density plane. 
Simultaneously varying $X$ and $\mu$ therefore defines a ``pQCD window'' that is equivalent to spanning the envelope of Fig.~\ref{fig:pqcd_uncertainty}. 
We plot this window in Fig.~\ref{fig:pqcd_window}, which can be compared with the lower panel of Fig.~\ref{fig:pqcd_uncertainty}. 
We choose to vary $X$ from 1/2 (lower red line) to 2 (upper blue line), and the chemical potential from 2.4 to $3\,\mathrm{GeV}$.
We consider any EOS that crosses this window to be consistent with the pQCD limit.
Effectively, instead of imposing that EOSs are consistent with pQCD at a fixed value of the chemical potential, we use the less-stringent requirement that an EOS can be consistent with pQCD at any chemical potential above $2.4\,\mathrm{GeV}$ (in principle we could extend the window to arbitrarily high chemical potentials, but we find little gain beyond $3\,\mathrm{GeV}$).
The pQCD calculation becomes unreliable at lower chemical potential, and values lower than $2.4\,\mathrm{GeV}$ are not often used in the literature.
This approach assigns an equal weight to all values of $X$ between $1/2$ and 2, which differs from other works (for example, as discussed in Sec.~\ref{sec:maximized_method}, it is common to place a uniform prior on the log of $X$).

Since we are only keeping EOS draws which cross this window, and the window is a relatively small region in the space of possible EOS behaviors, our GP will suffer from a low efficiency. 
In order to maximize the efficiency whilst ensuring the GP is still flexible enough to explore a wide range of behaviors we revisit the means, kernel, and hyperparameters used in its construction.

Firstly, GP means that themselves do not satisfy the pQCD constraints, i.e. that do not hit the ``pQCD window'' in Fig.~\ref{fig:pqcd_window}, could result in a low efficiency.
Even for an EOS mean that satisfies pQCD, any subsequent conditioning (such as a stitch to smoothly attach to a crust) will in general make it inconsistent with pQCD.
To circumvent this, we pre-compute 1000 EOSs that have the correct low- and high-density behavior to use as means and impose no further conditioning.
This includes hitting the correct pressure, number density, and chemical potential at low and high pressure, and also the behavior that $\csq \rightarrow 1/3$ at high pressure.
These means were themselves computed via an ``intermediate'' GP, for which we give more details in Appendix~\ref{app:speed_of_sound_gp}.

\begin{figure*}
    \centering
    \includegraphics[width=\linewidth]{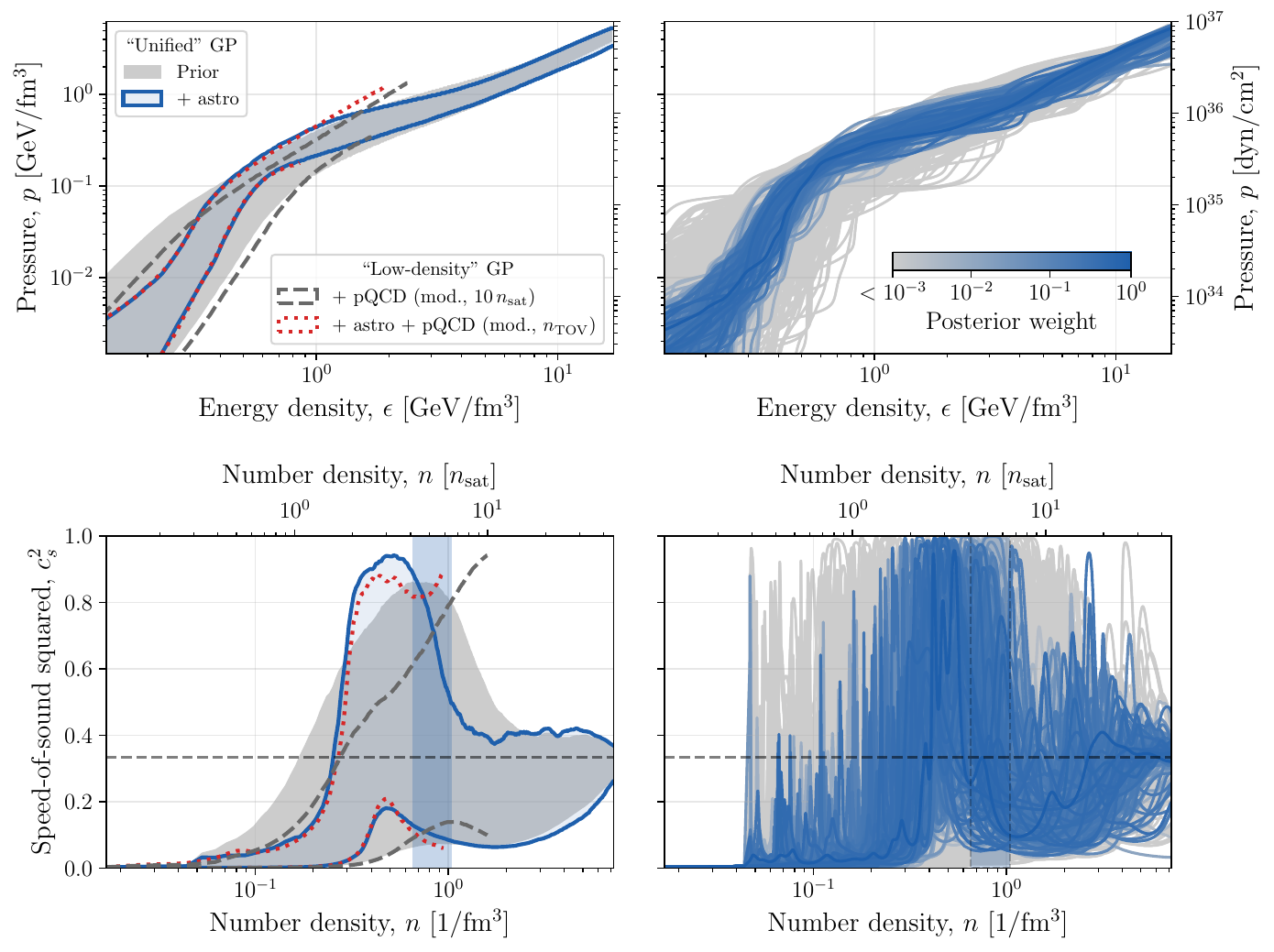}
    \caption{
    Prior (gray) and posterior when using all astronomical data (blue) for the unified GP.
    (Left panels) 90\% credible intervals.
    For comparison, we also plot in gray dashed the low-density GP posterior with the modeled pQCD constraint applied at $\nT=10\,\nsat$, see Fig.~\ref{fig:pqcd_posterior}, and in red dotted the posterior that additionally includes astronomical constraints, see Fig.~\ref{fig:astro_pqcd_posterior_gp0}.
    (Right panels) A sample of 1000 EOS draws from the unified GP colored by their posterior weight.
    The shaded vertical band on the speed-of-sound plots indicates the unified GP posterior on $\nTOV$(90\% credible interval).
    The posterior is very similar to the prior at densities above $\nTOV$, and even more so above $10\,\nsat$.
    }
    \label{fig:astro_pqcd_posterior_gp2}
\end{figure*}

Rather than performing any conditioning, we instead use a modified kernel which aims to follow the means more closely at low and high pressure and remain flexible at intermediate pressures.
To do this we generalize to a pressure-dependent kernel
\begin{align}
    \mathrm{Cov}[\phi(x), \phi(x')]
        & = k_\mathrm{SE}(x,x';\sigma,\ell) \nonumber \\
        & \quad\quad \times k_\mathrm{S}(x,x';\sigma_\mathrm{scale}^\mathrm{low},n_\mathrm{scale}^\mathrm{low}) \nonumber \\
        & \quad\quad \times k_\mathrm{S}(x,x';\sigma_\mathrm{scale}^\mathrm{high},n_\mathrm{scale}^\mathrm{high})\,,
\end{align}
where $x = \log[p/(1\,\mathrm{GeV/fm}{}^3)]$ are the GP test points where the EOS is evaluated.
This generalized kernel is a combination of the standard square-exponential kernel $k_\mathrm{SE}(x,x';\sigma,\ell)$ with hyperparameters $\sigma$ and $\ell$ drawn from the same distribution as Ref.~\cite{Essick:2019ldf} (see Appendix~\ref{app:gp2_flexibility} for details), and scaling factors
\begin{multline}
    k_\mathrm{S}(x,x';\sigma_\mathrm{scale},n_\mathrm{scale}) \\ = k_\mathrm{scale}(x;\sigma_\mathrm{scale},n_\mathrm{scale}) \times k_\mathrm{scale}(x';\sigma_\mathrm{scale},n_\mathrm{scale})\,,
\end{multline}
where
\begin{equation}
    k_\mathrm{scale}(x;\sigma_\mathrm{scale},n_\mathrm{scale}) = \sigma_\mathrm{scale} + \frac{1-\sigma_\mathrm{scale}}{1 + \left(x_\mathrm{ref}/x\right)^{n_\mathrm{scale}}}\,.
\end{equation}
By multiplying the $k_\mathrm{SE}$ kernel by two appropriate scaling kernels, we can force the GP variance to be smaller (and thus force draws to follow the mean more closely) at low and high pressures, while remaining very generic in between.
We select $\sigma_\mathrm{scale}^\mathrm{low} = 0$, and draw $\sigma_\mathrm{scale}^\mathrm{high} > 0$ from a truncated normal with mean 0 and standard deviation $10^{-2}$.
The remaining parameters $x_\mathrm{ref}^\mathrm{low}$, $x_\mathrm{ref}^\mathrm{high}$, $n_\mathrm{scale}^\mathrm{low}$, $n_\mathrm{scale}^\mathrm{high}$ are chosen to be normally distributed with means $\log(2 \times 10^{-3})$, $\log(0.9)$, 15, $-100$ and standard deviations 1, 0.5, 2.5, 15 respectively.
We draw 1000 hyperparameter realizations, which along with the means result in 1000 GP realizations. 
The final GP prior consists of an equal number of draws from each GP realization.
From this ``unified'' GP we obtain 53,911 pQCD-consistent EOSs from $6.5 \times 10^6$ total draws, for an efficiency of $\sim 0.83\%$.

\subsection{Unified GP results}
\label{sec:unified_results}

The prior for the unified GP and posterior when combined with astrophysical constraints are shown in Fig.~\ref{fig:astro_pqcd_posterior_gp2}, which can be directly compared with the low-density GP posteriors in Fig.~\ref{fig:astro_pqcd_posterior_gp0}.
The prior, by construction, satisfies pQCD constraints and so it is noticeably tighter compared to the prior of the low-density GP.
For comparison, we show posteriors for the low-density GP when only pQCD is applied (gray dashed line, which can be compared with the unified GP prior), and when both astrophysical and pQCD constraints are applied (red dotted, which can be compared with the unified GP posterior).
When applying only pQCD to the low-density GP we apply the constraints at a fixed number density, $\nT=10\,\nsat$ (as in Figs.~\ref{fig:pqcd_posterior} and \ref{fig:pqcd_posterior_nterm}) instead of $\nT=\nTOV$ because of a large number of EOSs in the prior with low values of $\nTOV$ (which makes plotting truncated quantiles in the pressure vs energy density and speed of sound vs number density space unfeasible). 

\begin{figure*}
    \centering
    \includegraphics[width=\linewidth]{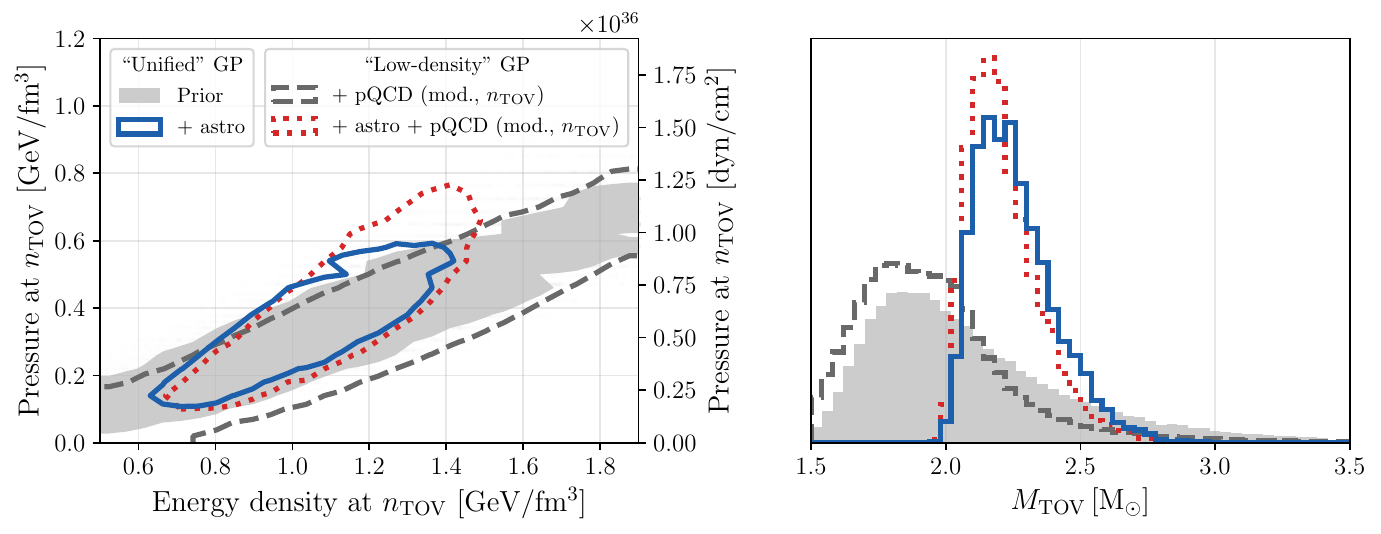}
    \caption{
    Pressure and energy density at $\nTOV$ (left) and $\MTOV$ (right) when applying astrophysical constraints to the unified GP.
    In each case we show the prior that automatically satisfies pQCD (gray) and the posterior (blue).
    As in Fig.~\ref{fig:astro_pqcd_posterior_gp2}, we also show constraints from applying only pQCD constraints to the low-density GP (dashed gray, which should be compared with the new prior), and constraints from applying astrophysical constraints alongside pQCD (dotted red, which should be compared with the new posterior).
    }
    \label{fig:p_eps_M_tov_gp2}
\end{figure*}

Comparing the unified GP prior to the gray dashed lines in Fig.~\ref{fig:astro_pqcd_posterior_gp2} (both corresponding to the EOS model + pQCD), there is a slight shift in the pressure vs energy density at the 90\% level---possibly a result of comparing constraints applied at $10\,\nsat$ instead of $\nTOV$.
Still, the unified GP is exploring a wide range of prior behaviors (a good indicator that the unified GP is not too restricted).
At a given energy density the unified GP is reaching higher pressures than the low-density GP + pQCD, and in the bottom left panel the unified GP is reaching higher sound speeds.
When astrophysical constraints are incorporated, the agreement is much better at densities $\lesssim 2\,\nsat$ between the unified and low-density GPSs (blue and red dashed, respectively), but as we approach central densities we see first that the unified GP posterior is slightly stiffer than the low-density GP + pQCD (this is more visible in the speed of sound squared posterior, where the upper bound on the speed of sound squared is higher for the unified GP) and then at $\nTOV$ the unified GP becomes softer than the low-density GP + pQCD.
Note that this softening may be related to the fact that we no longer see a widening of the speed-of-sound quantiles at the termination density (in this case, $\nTOV$), which may come from a weakening of the constraints near the termination density (see the discussion of Fig.~\ref{fig:astro_pqcd_posterior_gp0} in Sec.~\ref{sec:synergy}).
The difference in these posteriors demonstrates how the choice of the EOS behavior from NS central densities to pQCD densities can impact the constraining power of pQCD; the unified GP is built with different (more flexible) hyperparameters than the modeled pQCD likelihood, and the exact meaning of being consistent with pQCD is also different (our pQCD window is a looser requirement in the sense that we allow the pQCD constraint to be applied at any chemical potential $> 2.4\,\mathrm{GeV}$).

\begin{figure*}
    \centering
    \includegraphics[width=\linewidth]{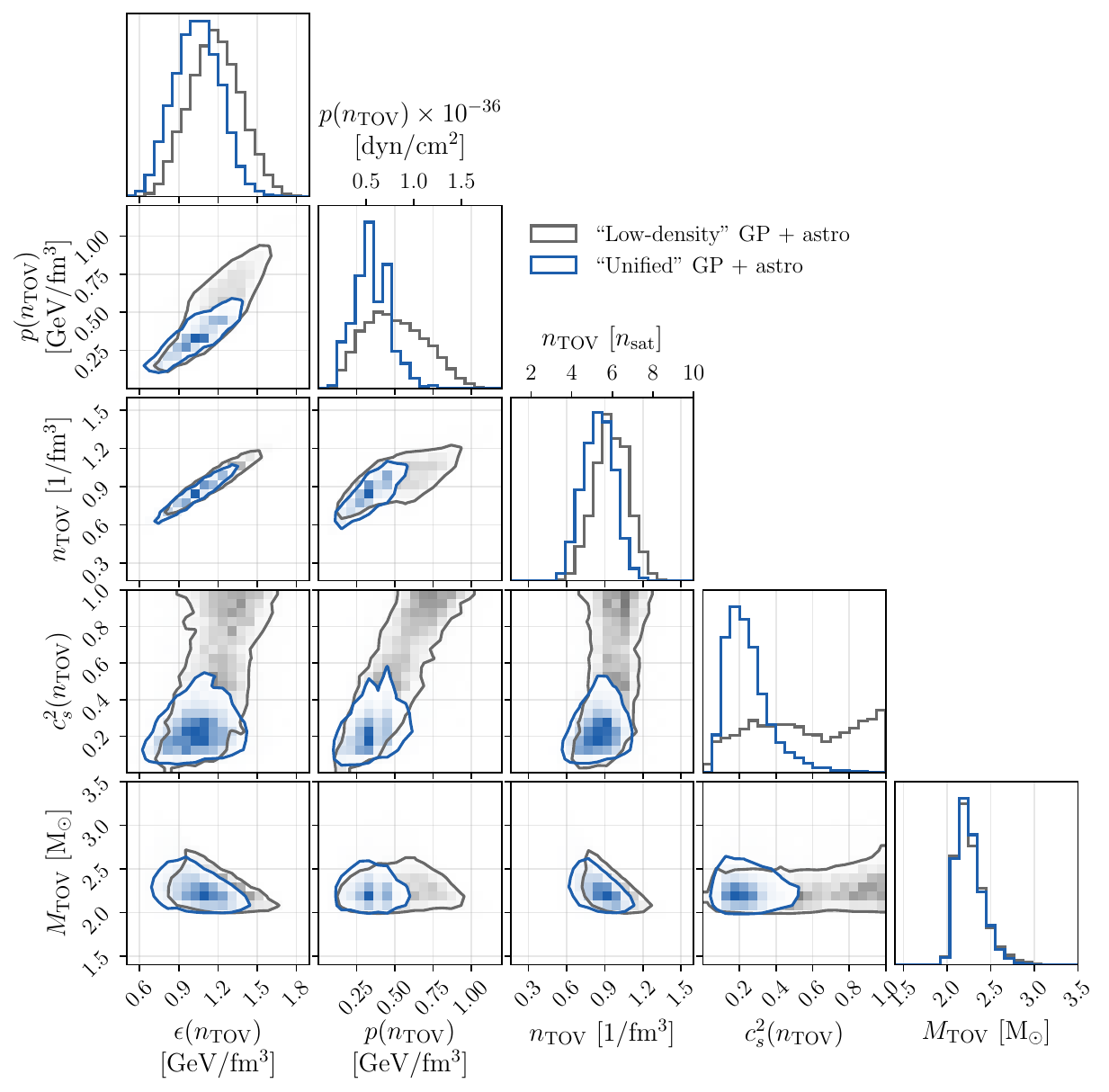}
    \caption{
    Posteriors on the TOV quantities (90\% credible regions) when applying all astrophysical constraints to the unified GP. 
    For comparison, we show posteriors from the low-density GP which also has astrophysical constraints applied, but no pQCD input.
    The differences in the posteriors show the impact of the pQCD; the most notable differences being a reduced upper limit on the pressure and speed of sound at $\nTOV$. 
    }
    \label{fig:tov_corner}
\end{figure*}

The right panels of Fig.~\ref{fig:astro_pqcd_posterior_gp2} show individual EOS draws colored by their posterior weight. 
The speed-of-sound plot shows that many draws reach the causal limit of $\csq=1$, but cannot sustain such a large speed of sound across densities as they would then overshoot the pQCD pressure.
In fact, every EOS in the unified GP posterior has a maximum speed of sound squared greater than $1/3$.
This is in agreement, but statistically stronger, than previous results indicating that the conformal limit of $1/3$ is indeed violated in NSs~\cite{Bedaque:2014sqa, Alsing:2017bbc, Tews:2018kmu, McLerran:2018hbz, Reed:2019ezm, Legred:2021hdx}.
After reaching its maximum, the speed of sound sharply drops around $\sim \nTOV$, with an upper limit of $\sim 0.4$ for $\csq$ at 90\% credibility for densities beyond $10\,\nsat$.
This drop in the speed of sound at TOV densities is one of the key features of the unified GP, and is discussed further below alongside Fig.~\ref{fig:tov_corner}.

We continue exploring the impact of how pQCD constraints are incorporated in Fig.~\ref{fig:p_eps_M_tov_gp2} with the pressure and energy density at $\nTOV$, and the TOV mass.
As in Fig.~\ref{fig:astro_pqcd_posterior_gp2}, we compare posteriors from the unified GP against posteriors from the low-density GP when applying pQCD via the likelihood; here we use the ``modeled'' likelihood for both the pQCD-only (gray dashed contour) and astro + pQCD (red dotted contour) posteriors.
For the pressure at $\nTOV$, we see that some of the highest values are now ruled out by the unified GP; this results in a slightly stronger constraint of $p(\nTOV) = 0.34^{+0.21}_{-0.16}\,\mathrm{GeV/fm}^3$ for the unified GP with astrophysical constraints, as compared to $0.36^{+0.33}_{-0.19}\,\mathrm{GeV/fm}^3$ for the low-density GP with pQCD and astrophysical constraints applied.
This is likely due to the aforementioned stiffening seen at the termination density whenever the pQCD constraints are applied via a likelihood; in the left-hand panels of Fig.~\ref{fig:astro_pqcd_posterior_gp2} we see this stiffening as a rise in the pressure and speed of sound squared around $\nTOV$ for the low-density GP with astrophysical constraints and pQCD applied (red dotted), but this is not seen in the unified GP (solid blue) since the pQCD constraints become part of the prior, so there is no termination density and so no artifacts.
Despite the stronger pressure constraint, we actually see a slight broadening in the $\MTOV$ posterior, where for the unified GP we have $\MTOV = 2.24^{+0.31}_{-0.17}\,\mathrm{M}_\odot$, as opposed to $\MTOV = 2.19^{+0.29}_{-0.15}\,\mathrm{M}_\odot$ for the low-density GP with pQCD and astrophysical constraints applied.
These differences express the quantitative impact of various modeling choices in applying the pQCD constraints. 

Finally, we address the key question of what is the actual impact of pQCD on NS densities in Fig.~\ref{fig:tov_corner}.
Since pQCD's impact is expected to be strongest for higher densities, we present results for select TOV quantities and specifically the unified GP and low density GP posteriors with all astronomical data, but with no pQCD constraints on the latter.
The difference between the two sets of posteriors, therefore, is indicative of the impact of pQCD.
The main impact is a reduction in all quantities other than $\MTOV$ that remains unchanged contrary to results reported elsewhere~\cite{Gorda:2022jvk, Koehn:2024set}.
The clearest differences are a $\sim 30\%$ reduction in the upper limit of the pressure at $\nTOV$, and a upper bound on the speed of sound squared at $\nTOV$ of $\sim 0.5$, which was previously unbounded.
The energy density shows a similar uncertainty for the unified GP but a slight shift to smaller values, going from $\epsilon(\nTOV) = 1.18^{+0.33}_{-0.31}\,\mathrm{GeV/fm}^3$ for the low density GP to $\epsilon(\nTOV) = 1.04^{+0.29}_{-0.28}\,\mathrm{GeV/fm}^3$ for the unified GP.
The same is true for the number density, going from $\nTOV = 5.94^{+1.30}_{-1.36}\,\nsat$ for the low-density GP to $\nTOV = 5.36^{+1.18}_{-1.26}\,\nsat$ for the unified GP.
Though not plotted here, we have verified that other observables (namely, the radius and tidal deformability at 1.4 and 2 solar masses) are all unchanged and (at least under the unified GP) pQCD has no observable impact on NS properties.
The lack of impact on astrophysical observables, despite the softening seen in the thermodynamic quantities, can be traced to a slight stiffening of the EOS at low densities; in the bottom left panel of Fig.~\ref{fig:astro_pqcd_posterior_gp2} we see that the unified GP exhibits a slightly steeper initial rise in the speed of sound squared compared to the low-density GP with astro + pQCD (which itself rises at essentially the same rate as the low-density GP with only astrophysical constraints---see Fig.~\ref{fig:astro_pqcd_posterior_gp0}). 
Since the TOV mass and radius depend on an integral over the pressure across all densities, the softening at TOV densities is effectively offset by the low-density stiffening.
And in Fig.~\ref{fig:tov_corner}, to gauge the impact of pQCD, we are comparing with a low-density GP which has a particular choice of $\MTOV$ prior, and so our comparison between the unified GP and the low-density GP is affected by that prior choice.
Since the unified GP incorporates pQCD into the prior by construction, which will change the $\MTOV$ prior, a comparison between a posterior without pQCD and one with pQCD (with the same $\MTOV$ prior for both) is not straightforward.

\section{Discussion and Conclusions}
\label{sec:discussion}

Calculations from pQCD offer information on the high-density EOS, beyond densities found in the cores of NSs.
However, based on arguments of thermodynamic stability and causality, pQCD constraints can be propagated to lower densities that are applicable to NSs.
In this work we verified previous results that apply the pQCD constraints via a likelihood~\cite{Gorda:2022jvk, Komoltsev:2023zor, Koehn:2024set}, using a very flexible EOS prior~\cite{Landry:2018prl, Essick:2019ldf, Landry:2020vaw}.
We confirmed that the impact of pQCD is to cause a softening of the EOS at the highest NS densities, with a small impact on macroscopic observables.
We highlighted how choices in the extrapolation procedure (referred to as a ``maximized'' and ``modeled'' likelihood) impact posterior constraints; we confirm that the ``maximized'' likelihood offers little constraint when applied at $\nTOV$ (but can provide constraints when applied at higher densities---see Figs.~\ref{fig:pqcd_posterior_nterm} and \ref{fig:pqcd_astro_e_p_tov_posterior}), whereas the ``modeled'' likelihood is less sensitive to the termination density (particularly when applied alongside astrophysical constraints) and shows a stronger impact.

Conceptually, however, both methods rely on an ad hoc choice for a termination density at which the allowed EOS behavior changes.
We avoid this switching of EOS representations by constructing a ``unified'' EOS model spanning the entire density range from the NS crust up to where pQCD is valid (akin to, for example, Refs.~\cite{Kurkela:2014vha, Fraga:2015xha, Annala:2017llu, Annala:2019puf, Annala:2021gom, Altiparmak:2022bke, Shirke:2022tta, Semposki:2024vnp, Semposki:2025etb}).
By incorporating pQCD into the EOS prior, this approach also treats knowledge from pQCD in a similar way to low-density knowledge (for example, from $\chi$EFT).
It is unified in the sense that we have a single description for the EOS across all densities of interest, whereas applying the pQCD constraints via a likelihood can be viewed as switching between different models at the termination density.
Given that the EOS from $\nTOV$ to the densities relevant for pQCD is currently unconstrained by any calculation or data, we also aim to maximize the flexibility of the unified EOS model in those densities.
We do this by employing a GP to interpolate between the two density regimes.
Still, the unified GP is constructed such that the EOS approaches the pQCD limit ``smoothly,'' see Fig.~\ref{fig:unified_mu_n}, and thus extreme behavior such as the one allowed by the maximized likelihood is statistically disfavored.

We found quantitatively similar results to applying the modeled likelihood to a low-density EOS model; i.e., that pQCD results in a softening of the EOS close to the highest densities encountered in NSs.
Most strikingly the central speed of sound squared of the maximum-mass NS is constrained to drop below $\sim 0.5$ after an initial sharp increase at densities $\lesssim 1.5-2\,\nsat$. 
Though the sharp increase above the conformal limit of $1/3$ inside NSs is well established based on existing astronomical data, the subsequent drop is a direct consequence of pQCD.
This is a combination of two facts. 
Firstly, asymptotic freedom implies that $\csq\rightarrow1/3$ as the density goes to infinity.
Secondly, the perturbative expansion about infinite density results in a value for the pressure of matter at high densities. 
These two facts combined imply not only that the speed of sound needs to decrease after its initial sharp rise, but also that it cannot remain large over a wide range of densities, since then the EOS would overshoot the pQCD pressure (as the speed of sound is related to the derivative of the pressure).
The exact quantitative picture of where and how the speed of sound decreases likely depends on prior assumptions and modeling choices at high densities.
However, the overall quantitative outcome remains, namely that the speed of sound needs to start decreasing inside the cores of NSs.

Perhaps disappointingly, with the unified GP, we find no impact on macroscopic observables, such as NS masses, radii, and tidal deformabilities.
Since there was an impact on macroscopic observables under the modeled pQCD likelihood, this highlights further that the method of extending the EOS from $\nTOV$ to pQCD densities plays an important role in how constraining pQCD is at NS densities.
It also suggests that astronomical observations themselves will offer little input on the properties of dense matter in pQCD densities.

Both the impact we see on $\csq$ and (lack thereof) on macroscopic properties at NS densities from pQCD are predicated on the assumptions that go into the construction of the unified GP.
Firstly, as is common with GPs, the choice of kernel and hyperparameters governs the behavior of the EOS draws and plays a role in the result. 
Indeed, this may be the critical difference between the unified GP and the EOS extensions that constitute the modeled likelihood.
Secondly, we expect our result to depend on the criteria for an EOS being consistent with pQCD when constructing the unified GP prior; we defined a ``pQCD window'' which we required our EOSs to pass through.
In defining this window we chose to vary the dimensionless renormalization scale $X$ between $1/2$ and 2 (a common choice in the literature), and varied the pQCD baryon chemical potential $\muH$ between 2.4 and $3\,\mathrm{GeV}$.
The lower limit of this range pushes the pQCD calculation to a region of large uncertainty (see the top panel of Fig.~\ref{fig:pqcd_uncertainty}), and was chosen to maximize efficiency of the unified GP. 
How choices for the window (or incorporating pQCD uncertainties, as in \citet{Gorda:2023usm}) impact the result would be an interesting direction to explore.
And thirdly, in constructing the unified GP means, we made a choice for the length scale that we transition to $\csq = 1/3$ as a function of pressure (see Appendix~\ref{app:speed_of_sound_gp}).
We modeled this transition on the EOS extensions of \citet{Komoltsev:2023zor}, which start tending toward the pQCD limit at as low as $\sim 25\,\nsat$.
It is expected that this choice is the dominant contribution to the reduced speed of sound seen at TOV densities, and it would be interesting to quantify how the impact of pQCD varies as a function of this transition length scale to $\csq = 1/3$.
All these extensions as well as aiming toward more model flexibility will come to the forefront as astronomical data increase in constraining power and the question of pQCD impact is revisited.

Draws from the unified GP as well as astrophysical weights are available at Ref.~\cite{finch_2025_15390534}.
The code used to perform all the analyses and create all figures is available at Ref.~\cite{pqcd}.


\begin{acknowledgments}

We thank Marco Crisostomi and Cole Miller for helpful discussions about renormalization in QCD and the relationship between thermodynamic EOS quantities and astrophysical observables.
E.F., I.L., and K.C. acknowledge support from the Department of Energy under award number DE-SC0023101, the Sloan Foundation, and by a grant from the Simons Foundation (MP-SCMPS-00001470).
R.E. is supported by the Natural Sciences \& Engineering Research Council of Canada (NSERC) through a Discovery Grant (RGPIN-2023-03346). 
The work of S.H. was supported by Startup Funds from the T.D. Lee Institute and Shanghai Jiao Tong University. 
The authors are grateful for computational resources provided by the LIGO Laboratory and supported by National Science Foundation Grants PHY-0757058 and PHY-0823459.
This material is based upon work supported by NSF's LIGO Laboratory which is a major facility fully funded by the National Science Foundation.

\end{acknowledgments}


\appendix

\begin{figure}[b]
    \centering
    \includegraphics[width=\columnwidth]{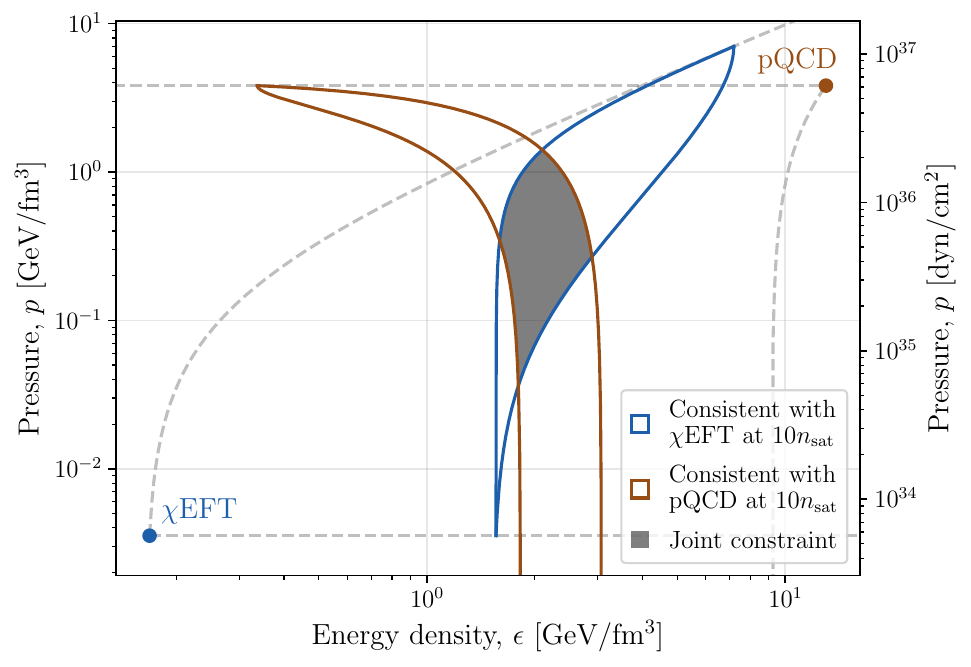}
    \caption{
    When pQCD is considered jointly with $\chi$EFT, we can get tighter constraints at a particular density (see Fig.~1 in Ref.~\cite{Komoltsev:2021jzg}).
    Given that the EOS must pass through a low-density $\chi$EFT prediction (blue marker), then at some specified density (taken here to be $10\nsat$) thermodynamic consistency/stability and causality define an allowed region (blue line).
    Likewise, a pQCD prediction also defines an allowed region (this is the ``maximized'' pQCD likelihood, see the left panel of Fig.~\ref{fig:max_vs_marg}).
    The overlap of these regions gives the joint constraint (dark gray area).
    For the $\chi$EFT values we take the ``stiff'' EOS at $1.1 \,\nsat$ from Ref.~\cite{Hebeler:2013nza}.
    For the pQCD values we take $\muH = 2.6\,\mathrm{GeV}/\mathrm{fm}^3$ and $X=1$.
    Dashed lines show the paths of EOSs with $\csq = 0$ and 1 which intersect the $\chi$EFT and pQCD values.
    }
    \label{fig:joint_constraint}
\end{figure}

\section{Allowed region in the energy density -- pressure plane}
\label{app:epsilon-p}

\begin{figure}
    \centering
    \includegraphics[width=\columnwidth]{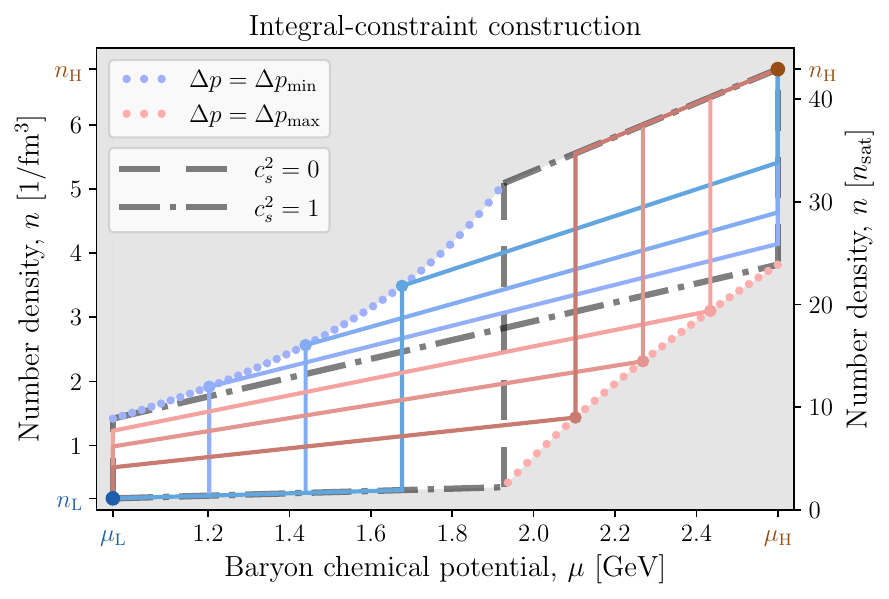}
    \caption{
    Given that an EOS must pass through a point at low density, $(\muL,\nL,\pL)$ (which could come from, for example, a $\chi$EFT prediction), and a point at high density, $(\muH,\nH,\pH)$ (for example, a pQCD prediction), certain regions of the EOS phase space can be ruled out (light gray region) based only on thermodynamic consistency/stability, and causality. 
    Blue lines are EOSs that minimize the pressure change whilst passing through a particular point in the phase space (blue dots), and red lines maximize the pressure change whilst passing through a particular point (red dots); the set of blue and red markers are the points for which the minimum and maximum pressure changes equal the required pressure change $\Delta p = \pH - \pL$. 
    For the $\chi$EFT values we take the ``soft'' EOS at $1.1 \,\nsat$ from Ref.~\cite{Hebeler:2013nza}. For the pQCD values we take $\muH = 2.6\,\mathrm{GeV}/\mathrm{fm}^3$ and $X=2$.
    An EOS that goes above the blue markers would always accumulate too much pressure to hit the pQCD value, and an EOS that goes below the red markers would never accumulate enough pressure. 
    }
    \label{fig:constraint_construction}
\end{figure}

In Fig.~\ref{fig:joint_constraint} we show how simultaneous consistency with low-density $\chi$EFT values and high-density pQCD values define a region in the $\epsilon$--$p$ plane at fixed number density; this can be compared with Fig.~1 in Ref.~\cite{Komoltsev:2021jzg}.
Although not directly used in the analysis, this figure clearly illustrates how the pQCD constraints presented in Ref.~\cite{Komoltsev:2021jzg} are constructed and how they are related to the ``maximized'' pQCD likelihood (left panel of Fig.~\ref{fig:max_vs_marg}).
Given that the EOS has to pass through some point, $(\muL,\nL,\pL)$, then at a specified density $n_0$ (in the figure chosen to be $10\,\nsat$) there is a limited region allowed in the $\epsilon$--$p$ plane based only on thermodynamic consistency/stability, and causality (blue line).
The construction of this region is most easily done in the $\mu$--$n$ plane.
First, at $n_0$ the corresponding $\mu_0$ is bounded between $\muL$ (given by an EOS with $\csq = 0$ from $\nL$ to $n_0$) and $(n_0/\nL)\muL$ (given by an EOS with $\csq = 1$ from $\nL$ to $n_0$). 
Then for each $\mu_0$ we can obtain lower and upper bounds on the corresponding $p_0$ by considering EOSs that minimize and maximize the pressure change from $\muL$ to $\mu_0$ (using the same idea as in Fig.~\ref{fig:constraint_construction}). 
Finally, we obtain the corresponding $\epsilon_0$ for the lower and upper pressure bounds via $\epsilon = \mu n - p$.
We use the same logic to obtain the red region in Fig.~\ref{fig:joint_constraint}, except that we work backward from $(\muH,\nH,\pH)$ as given by pQCD.
Now, at $n_0$, $\mu_0$ is bounded between $(n_0/\nH)\muH$ and $\muH$.
Where these regions overlap gives us the allowed region at $n_0$.

\begin{figure*}
    \centering
    \includegraphics[width=0.7\linewidth]{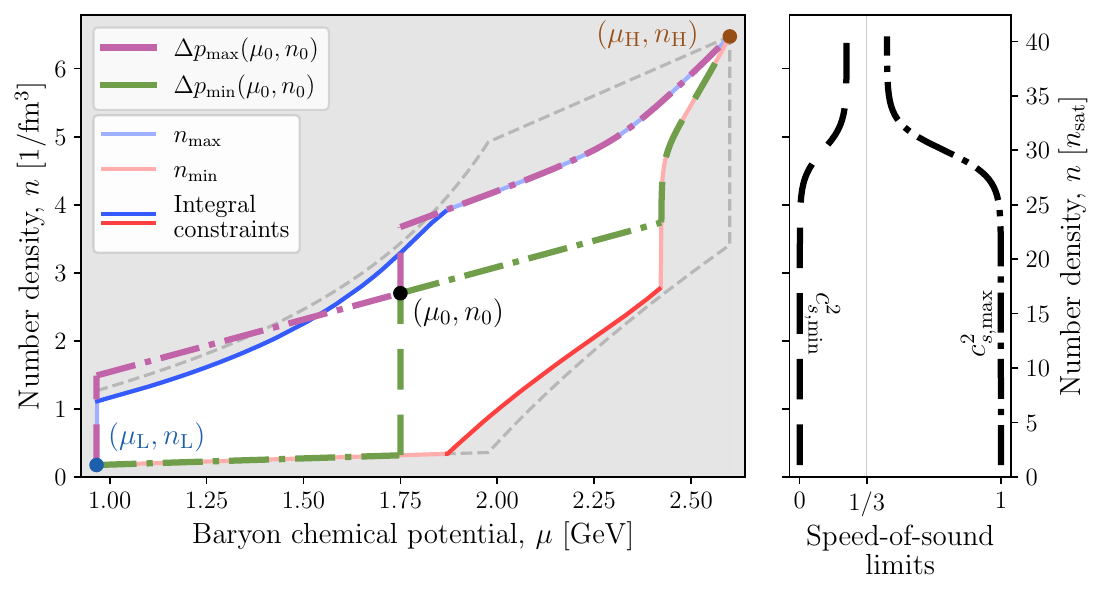}
    \caption{
    Based on a similar figure in Ref.~\cite{Komoltsev:2021jzg}, here we show the ruled-out region of the $\mu$--$n$ space (left panel) when we impose speed-of-sound limits that vary as a function of $n$ (right panel).
    To construct the integral constraints, we must be able to find the $\Delta p_\mathrm{min/max}(\mu_0,n_0)$ curves that connect $(\muL,\nL)$ and $(\muH,\nH)$ whilst passing through an arbitrary $(\mu_0,n_0)$ (examples of such curves are shown in purple and green).
    The loci of $(\mu_0,n_0)$ points for which $\Delta p_\mathrm{min/max} = \pH - \pL$ define the ``integral constraints,'' which, together with the causal constraints, constitute the $n_\mathrm{max/min}$ boundary.
    Shown in the dashed gray line on the left panel is the allowed region with $c_{s,\mathrm{min}}^2 = 0$ and $c_{s,\mathrm{max}}^2 = 1$ (as also shown in Fig.~\ref{fig:constraint_construction}).
    }
\label{fig:general_cs2_limits}
\end{figure*}

\section{pQCD integral constraints with generalized speed-of-sound limits}
\label{app:cs2}

In this appendix we illustrate how the ``integral'' constraints of Ref.~\cite{Komoltsev:2021jzg} are constructed, and how they can be generalized for generic prescriptions of the speed of sound as a function of density.
Although not used in our analysis, this discussion helps to clarify the integral constraints and could be useful for constructing EOS models with the pQCD constraints built in.

In Fig.~\ref{fig:constraint_construction} we show how the constraints presented in Ref.~\cite{Komoltsev:2021jzg} are constructed.
At low density the EOS is under good theoretical control via $\chi\mathrm{EFT}$, and we plot an example prediction for the thermodynamic quantities $(\muL,\nL,\pL)$.
Specifically, here we take values from the ``soft'' EOS of Ref.~\cite{Hebeler:2013nza} at $1.1\, \nsat$.
For a choice of chemical potential $\muH$ and dimensionless renormalization scale $X$ (taken to be $2.6\,\mathrm{GeV}$ and $2$ in Fig.~\ref{fig:constraint_construction}, respectively), pQCD gives us a prediction of the EOS at high density, $(\muH,\nH,\pH)$.
Both of these predictions have uncertainties associated with them, and so any analysis should marginalize over this uncertainty. 
Here, for illustrative purposes, we fix the low- and high-density points, so we have two points that our EOS must pass through.

Given these constraints, it is possible to rule out certain regions of the $\mu$--$n$ plane for which no stable/consistent/causal EOS can connect $(\muL,\nL,\pL)$ and $(\muH,\nH,\pH)$ while accumulating the correct pressure.
The boundaries of the allowed region are found by considering EOSs that, given they must pass through a particular point in the $\mu$--$n$ plane, either minimize or maximize the pressure change between $(\muL,\nL)$ and $(\muH,\nH)$ (blue and red lines in the figure, respectively). 
The loci of points for which the minimum pressure change $\Delta p_\mathrm{min}$ equals the required pressure change $\Delta p$ constitute the boundary $n_\mathrm{max}(\mu)$ (blue markers), and the loci of points for which the maximum pressure change $\Delta p_\mathrm{max} = \Delta p$ constitute $n_\mathrm{min}(\mu)$ (red markers).
Explicit expressions for these boundaries are given in Ref.~\cite{Komoltsev:2021jzg}.

These integral constraints derived in Ref.~\cite{Komoltsev:2021jzg} are based on a fixed value for the maximum speed of sound squared; that is, the maximum (or limiting) speed of sound squared (denoted $c_{s,\mathrm{lim}}^2$ in that work) was not allowed to vary as a function of the number density (in Fig.~\ref{fig:constraint_construction}, we impose the most conservative causal limit $c_{s,\mathrm{lim}}^2 = 1$).
In addition, the influence of the minimum $\csq$ on the integral constraints was not considered.
However, at high density the speed of sound squared should approach the conformal limit of $\csq = 1/3$. 
Here we show how to incorporate this knowledge into the integral constraints, which further shrinks the available space that candidate EOSs can occupy.

In order to obtain the integral constraints, we must be able to compute $\Delta p_\mathrm{min}(\mu_0,n_0)$ and $\Delta p_\mathrm{max}(\mu_0,n_0)$.
That is, we must be able to find the paths from $(\muL,\nL)$ to $(\muH,\nH)$, passing through a given point $(\mu_0,n_0)$, which accumulate the maximum and minimum possible pressures.
Then, the set of $(\mu_0,n_0)$ points for which $\Delta p_\mathrm{min}(\mu_0,n_0) = \pH - \pL = \Delta p$ defines the curve $n_\mathrm{max}(\mu)$, and the set of points for which $\Delta p_\mathrm{max}(\mu_0,n_0) = \Delta p$ define $n_\mathrm{min}(\mu)$.
When the speed-of-sound limits were of the form $\csq < c_{s,\mathrm{lim}}^2 = \mathrm{constant}$, analytic expressions for $\Delta p_\mathrm{min/max}(\mu_0,n_0)$ and $n_\mathrm{max/min}(\mu)$ could be written down.
However, we now want to apply more generic, density-dependent speed-of-sound limits $c_{s,\mathrm{min/max}}^2(n)$ that capture the knowledge $\csq \rightarrow 1/3$ at high density.
In Fig.~\ref{fig:general_cs2_limits} we show a simple example where we have used sigmoid functions to define an envelope of allowed $\csq$ values as a function of number density.
The right panel shows $c_{s,\mathrm{min/max}}^2(n)$, which at low density are agnostic, but at high density only allow a limited range of values around $\csq = 1/3$.
The left panel shows, for a specific $(\mu_0,n_0)$, the paths which accumulate the maximum ($\Delta p_\mathrm{max}$, purple) and minimum ($\Delta p_\mathrm{min}$, green) pressures within the speed-of-sound constraints.
The computation of the $\Delta p_\mathrm{min/max}(\mu_0,n_0)$ curves will in general no longer be analytic.

To proceed, we write the (inverse of) speed of sound squared as
\begin{equation}
    c_s^{-2} = \frac{\mu}{n}\pdv{n}{\mu}\,.
\end{equation}
Setting $\csq = c_{s,\mathrm{min/max}}^2(n)$ and integrating we obtain
\begin{gather}
    \int \dd{n} \frac{c_{s,\mathrm{min/max}}^2(n)}{n} = I_\mathrm{min/max}(n) = \ln{\mu} + \mathrm{const}. 
\end{gather}
where $I_\mathrm{min/max}(n)$ is the (numerically computed) integral over $n$ with $\csq = c_{s,\mathrm{min}}^2(n)$ or $\csq = c_{s,\mathrm{max}}^2(n)$.
Requiring that we pass through $(\mu_0,n_0)$ sets the constant, giving
\begin{equation}
    \mu(n) = \mu_0 e^{I_\mathrm{min/max}(n) - I_\mathrm{min/max}(n_0)}\,.
\end{equation}

\begin{figure}[b]
    \centering
    \includegraphics[width=\columnwidth]{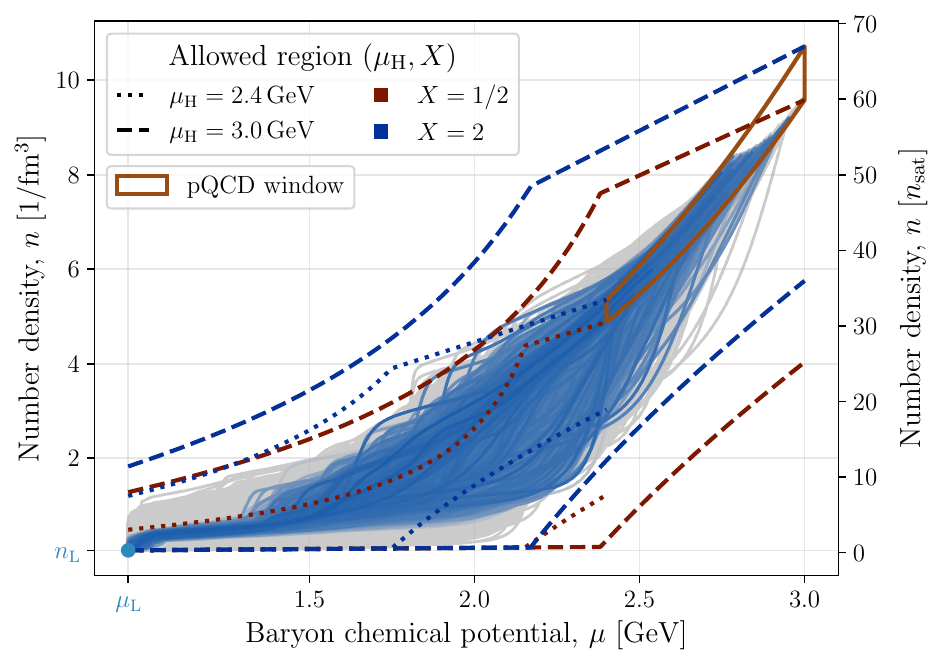}
    \caption{
    Sample of 25,000 unified GP EOSs plotted in the baryon chemical potential vs number density space (this is about half the total number of EOSs in our prior).
    As in Fig.~\ref{fig:astro_pqcd_posterior_gp2} we color the EOSs according to their posterior weight once all astrophysical constraints have been applied (the same color scale is used).
    We also indicate with the brown region the ``pQCD window'' of Fig.~\ref{fig:pqcd_window} which we require EOS draws to hit.
    Each EOS is terminated when it hits this window.
    The dotted and dashed lines indicate the allowed region in the $\mu$-$n$ plane given a choice of low- and high-density $(\mu,n,p)$, as in Figs.~\ref{fig:constraint_construction} and \ref{fig:general_cs2_limits} (see main text).
    }
    \label{fig:unified_mu_n}
\end{figure}

\begin{figure*}
    \centering
    \includegraphics[width=\linewidth]{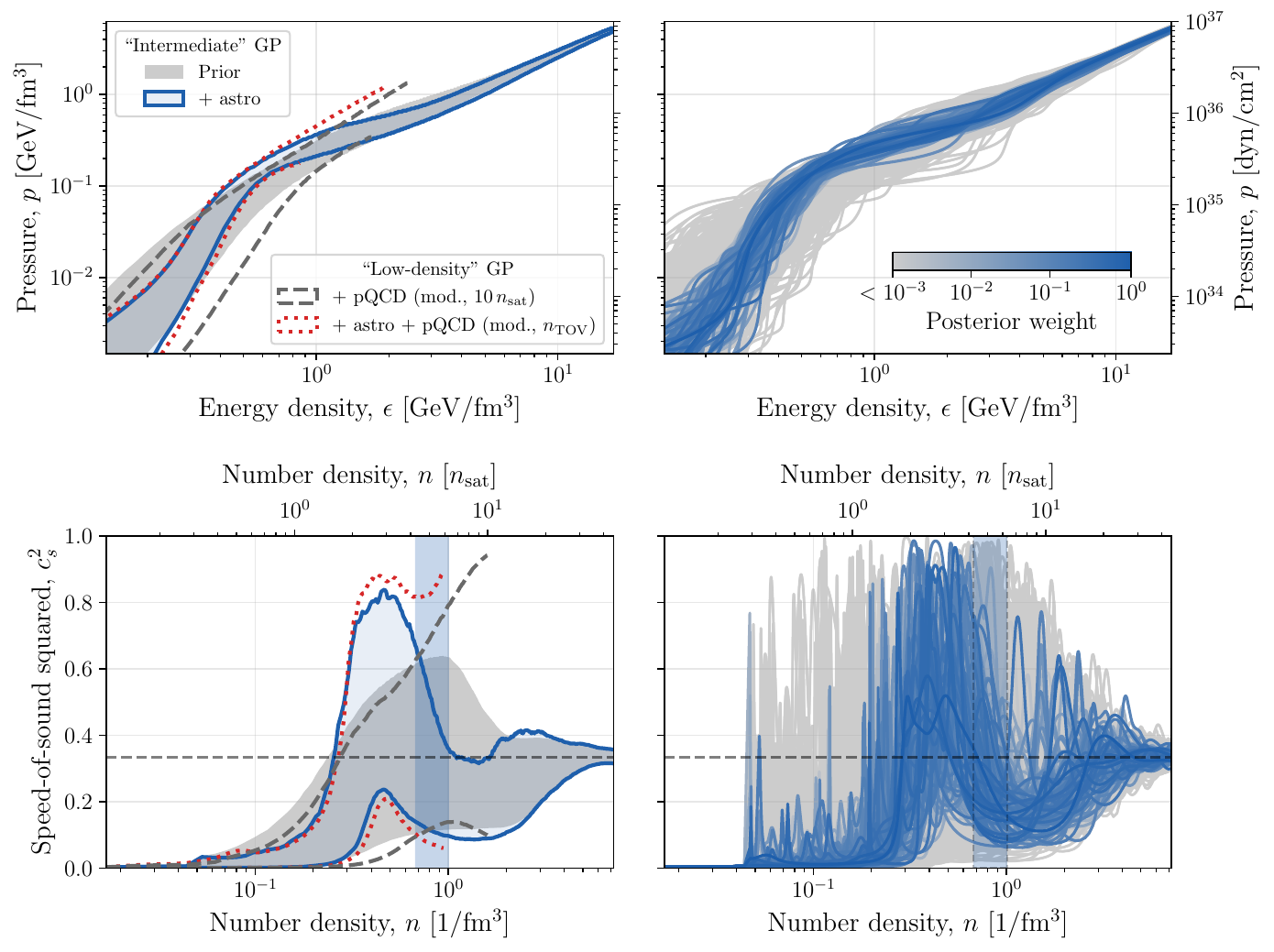}
    \caption{
    As in Fig.~\ref{fig:astro_pqcd_posterior_gp2}, but for the ``intermediate'' GP.
    }
    \label{fig:astro_pqcd_posterior_gp1}
\end{figure*}

This defines the paths in the $\mu$--$n$ plane that an EOS takes when $\csq = c_{s,\mathrm{min}}^2(n)$ or $\csq = c_{s,\mathrm{max}}^2(n)$.
The $\Delta p_\mathrm{min/max}(\mu_0,n_0)$ curves in Fig.~\ref{fig:general_cs2_limits} are constructed by considering the lines that minimize/maximize the area under the curves whilst passing through each of $(\muL,\nL)$, $(\mu_0,n_0)$ and $(\muH,\nH)$.

Finally, to find $n_\mathrm{max/min}(\mu)$, 
we iterate through each value of $\mu_0$ from $\muL$ to $\muH$ and at each step numerically find the pair of $n_0$ values for which $\Delta p_\mathrm{min/max} = \Delta p$.
The resulting curves are shown as blue and red lines in Fig.~\ref{fig:general_cs2_limits}.
The region outside these lines, shaded in light gray, is ruled out.
For comparison, the dashed gray lines show the boundaries of the previous integral constraints obtained with only the minimum, less restrictive requirement $\csq < 1$.

\section{Unified GP flexibility}
\label{app:gp2_flexibility}

In Fig.~\ref{fig:unified_mu_n} we plot a subset of unified EOS prior (gray) and posterior (blue) draws in the baryon chemical potential vs number density plane, and indicate the maximally allowed EOS region given low- and high-density points the EOS passes through (these allowed regions are based on thermodynamic consistency/stability, and causality; see Appendix~\ref{app:cs2} and Fig.~\ref{fig:constraint_construction}).
This allows us to gauge how much of the available phase space the EOS draws are exploring, but due to some caveats (explained below) this is an imperfect method for assessing model flexibility.

To construct the allowed regions (dotted and dashed lines), low-density $(\muL,\nL,\pL)$ and high-density $(\muH, \nH, \pH)$ points must be chosen. 
For the low-density $(\muL,\nL,\pL)$ we take the value where the EOS draws are stitched to a crust~\cite{Baym:1971pw, Hebeler:2013nza} (we stitch at a pressure of $0.168\,\mathrm{MeV} = 3\times10^{11}\,\mathrm{g/cm}^3$, and this is the same for all EOS draws). 
For the high-density values, any $(\muH,\nH,\pH)$ that lies within the pQCD window could be chosen. 
We indicate the window (which is the same as in Fig.~\ref{fig:pqcd_window}) with a brown box that extends from 2.4\,GeV to 3.0\,GeV in chemical potential; EOS draws terminate when they hit this window.
The biggest possible allowed region is obtained by taking $\muH = 3.0\,\mathrm{GeV}$ (dashed lines), and then varying the dimensionless renormalization scale $X$ (which determines $\nH$ and $\pH$) between $1/2$ (red) and 2 (blue).
Comparing the EOS prior draws (gray lines) to this region (blue and red dashed lines) assumes that all EOS draws hit the pQCD window at $3.0\,\mathrm{GeV}$.
We instead find that the majority of EOS draws hit the window at lower values of chemical potential, so this is not a fair representation of the available phase space.
The smallest allowed regions are obtained by taking $\muH = 2.4\,\mathrm{GeV}$ (dotted lines), and we again plot regions for $X = 1/2$ and $2$. 
EOS draws fill the space bounded by these lines, but this is also not a fair representation of the available space to the EOS draws.
The true flexibility of the unified GP is somewhere between these two extremes.

For completeness, here we quote the distributions for the additional hyperparameters used in the unified GP (which were inherited from the low-density GP~\cite{Essick:2019ldf}, see also Appendix A of Ref.~\cite{Legred:2022pyp}); these are $\ell \in U(0.1, 0.9)$ (where $\ell$ has units $\log(p/(\mathrm{g}/\mathrm{cm}^3))$, and $\log(\sigma) \in U(0, 1.6)$ (where $\sigma$ has units of $\phi$). 
Note, however, that since our GP is built on $\log(p)$, it is hard to make a direct comparison between the GP hyperparameters used here and those used in other works (for example, Eq.~(6) in Ref.~\cite{Gorda:2022jvk} or Eq.~(12) in Ref.~\cite{Komoltsev:2023zor}.

\section{Intermediate GP}
\label{app:speed_of_sound_gp}

The unified GP of Sec.~\ref{sec:pqcd_prior} used as means a sample of EOSs that were consistent with low-density (stitched to a NS crust) and high-density (intersect the pQCD window of Fig.~\ref{fig:pqcd_window}) behavior.
These means were generated via an ``intermediate'' GP, which was a simple extension of the low-density GP.
We give details on this intermediate GP in this appendix.

When designing a GP which extends to pQCD densities, the most straightforward extension is to retain the structure of the low-density GP, i.e., the tabulated EOSs upon which the process is conditioned, the kernel function and hyperparameters, and the crust stitching~\cite{Landry:2018prl, Essick:2019ldf, Landry:2020vaw}, and simply extend its test points (the points we evaluate the GP at) to pQCD pressures.
Since the GP is constructed in the $\phi$--$p$ space, we also condition the speed of sound (Eq.~(9) of Ref.~\cite{Essick:2019ldf}) so that $\csq \rightarrow 1/3$ at pQCD pressures (roughly matching the behavior seen in Fig.~10 of \citet{Komoltsev:2023zor}).
A large number of draws were made from this GP, and they went through the same filtering process as described in Sec.~\ref{sec:unified_construction}.
From $9 \times 10^6$ GP draws, we obtain 51,177 pQCD-consistent EOSs, for an efficiency of $\sim 0.57\%$.
A random sample of 1000 of these draws were used as means in the unified GP.

For comparison, Fig.~\ref{fig:astro_pqcd_posterior_gp1} shows the intermediate GP prior and posterior including astrophysical data.
The intermediate GP prior is noticeably narrower than the low-density GP with pQCD constraints applied, and the posterior with astrophysical constraints applied also explores less parameter space.
This is particularly clear in the speed of sound, where the blue intermediate GP quantiles do not reach as high a speed of sound as the low-density GP with astrophysical and pQCD constraints applied.
This motivates exploring more flexible models, which the unified GP of Sec.~\ref{sec:pqcd_prior} is designed to do.


\normalem
\bibliography{bibliography}

\end{document}